\documentclass[10pt,twocolumn,floatfix,prl,superscriptaddress]{revtex4-2}
\usepackage{amsmath,amssymb,amsthm,mathrsfs,amsfonts,dsfont,amstext}
\usepackage{textcomp,pbox}
\usepackage[export]{adjustbox}
\usepackage{bm}
\usepackage{dcolumn,booktabs,url}
\usepackage[scaled]{helvet}
\usepackage{sansmath,gensymb}
\usepackage{tikz,graphicx,transparent,color}
\usepackage{multirow}
\usepackage[separate-uncertainty = false]{siunitx}
\usepackage{comment}
\usepackage{physics}
\usepackage{pdfpages}
\usepackage{array}
\usepackage{xspace}

\makeatletter
\AtBeginDocument{\let\LS@rot\@undefined}
\renewcommand{\fnum@figure}{Fig. \thefigure} 
\makeatother
 
\newcolumntype{C}[1]{>{\centering\let\newline\\\arraybackslash\hspace{0pt}}m{#1}}

\usepackage[colorlinks=true]{hyperref}
\usepackage{graphicx}

\hypersetup{
     colorlinks   = true,
     citecolor    = blue,
     linkcolor    = blue,
     urlcolor     = blue     
}

\graphicspath{{./figs/}}
\newcommand{\down}{\downarrow}
\newcommand{\up}{\uparrow}
\newcommand{\Down}{\Downarrow}

\newcommand{\ketup}{{\ket{\uparrow}}}
\newcommand{\ketdown}{{\ket{\downarrow}}}

\newcommand{\fidelity}{68.6(4)\%\xspace}
\newcommand{\storagetime}{\SI{130(16)}{\micro s}\xspace}

\newcommand{\plusseventyone}{$^{71}$Ga$_{[+]}$\xspace}
\newcommand{\minusseventyone}{$^{71}$Ga$_{[-]}$\xspace}
\newcommand{\plusminusseventyone}{$^{71}$Ga$_{[\pm]}$\xspace}

\begin{document}

\newcommand{\TitleName}{Many-body quantum register for a spin qubit}

\title{\TitleName}
\newcommand{\AffCam}{Cavendish Laboratory, University of Cambridge, J.J. Thomson Avenue, Cambridge, CB3 0HE, UK}
\newcommand{\AffLinz}{Institute of Semiconductor and Solid State Physics, Johannes Kepler University, Altenberger Str. 69, Linz 4040, Austria}
\newcommand{\AffOxford}{Department of Engineering Science, University of Oxford, Oxford, UK}

\author{Martin Hayhurst Appel}
\affiliation{\AffCam}
\author{Alexander Ghorbal}
\affiliation{\AffCam}
\author{Noah Shofer}
\affiliation{\AffCam}
\author{Leon Zaporski}
\affiliation{\AffCam}

\author{Santanu Manna}
\affiliation{\AffLinz}
\author{Saimon Filipe Covre da Silva}
\affiliation{\AffLinz}

\author{Urs Haeusler}
\affiliation{\AffCam}
\author{Claire Le Gall}
\affiliation{\AffCam}
\author{Armando Rastelli}
\affiliation{\AffLinz}
\author{Dorian A. Gangloff}
\email[Correspondence to: ]{dag50@cam.ac.uk}
\affiliation{\AffCam}
\author{Mete Atat\"ure}
\email[Correspondence to: ]{ma424@cam.ac.uk}
\affiliation{\AffCam}

\date{\today}

\begin{abstract}
Quantum networks require quantum nodes with coherent optical interfaces and multiple stationary qubits. In terms of optical properties, semiconductor quantum dots are highly compelling, but their adoption as quantum nodes has been impaired by the lack of auxiliary qubits. Here, we demonstrate a functional quantum register in a semiconductor quantum dot leveraging the dense, always-present nuclear spin ensemble. We prepare 13,000 host nuclear spins into a single many-body dark state to operate as the register logic state $\ket{0}$. The logic state $\ket{1}$ is defined as a single nuclear magnon excitation, enabling controlled quantum-state transfer between the electron spin qubit and the nuclear magnonic register.
Using 130-ns SWAP gates, we implement a full write-store-retrieve-readout protocol with \fidelity raw overall fidelity and a storage time of \storagetime in the absence of dynamical decoupling. Our work establishes how many-body physics can add step-change functionality to quantum devices, in this case transforming quantum dots into multi-qubit quantum nodes with deterministic registers.
\end{abstract}

\maketitle 

Quantum nodes consisting of multiple qubits with efficient coupling to photons are required by a wide range of quantum information tasks including quantum repeaters~\cite{briegel_quantum_1998, wehner_quantum_2018, bhaskar_experimental_2020} and deterministic 2d-cluster state generation~\cite{buterakos_deterministic_2017, michaels_multidimensional_2021}. 
In one approach, an optically active spin qubit exchanges quantum information between a photonic mode and several long-lived register qubits ~\cite{pompili_realization_2021,bhaskar_experimental_2020}. 
Multiple spin-photon interfaces have demonstrated functional registers including diamond color centres coupled to proximal $^{13}$C nuclear spins~\cite{dutt_quantum_2007,taminiau_detection_2012,maurer_room-temperature_2012} or to the native nuclear spin of the color centre~\cite{stas_robust_2022}, SiC divacancy spins coupled to $^{29}$Si nuclear spins~\cite{bourassa_entanglement_2020}, multi-species ion traps~\cite{drmota_robust_2023}, and $^{171}$Yb$^{3+}$ ions coupled with neighboring $^{51}$V$^{5+}$ ions in a YVO$_4$ crystal \cite{ruskuc_nuclear_2022}. 
Group III-V semiconductor quantum dots (QDs) have state-of-the-art photon coherence \cite{zhai_quantum_2022,uppu_scalable_2020} and brightness \cite{tomm_bright_2021} but have so far lacked auxiliary qubits for the electron spin qubit. In contrast to the aforementioned few-particle systems, an electron spin qubit confined in a QD is Fermi-contact coupled to an ensemble of $\sim\!\!10^5$ nuclear spins \cite{urbaszek_nuclear_2013} which when uncontrolled acts as a source of noise for the qubit ~\cite{bechtold_three-stage_2015,
malinowski_notch_2016,stockill_quantum_2016}. Once sufficiently engineered, the ensemble can instead act as a bosonic system capable of encoding quantum information in collective excitations \cite{taylor_controlling_2003, taylor_long-lived_2003, ding_high-fidelity_2014, denning_collective_2019} similar to photon memories~\cite{kozhekin_quantum_2000}, ferromagnetic magnons ~\cite{tabuchi_coherent_2015} and microwave resonators~\cite{eickbusch_fast_2022}.

Significant progress on controlling dense nuclear spin ensemble using InGaAs QDs includes dynamical nuclear polarisation~\cite{hogele_dynamic_2012}, stabilization of the nuclear Overhauser field~\cite{gangloff_quantum_2019, jackson_optimal_2022}, and electron-mediated collective nuclear excitations \cite{jackson_quantum_2021}. Despite reaching low-fluctuation high-purity nuclear states, the coherence of the nuclear ensemble is however limited by the strain-induced nuclear broadening \cite{stockill_quantum_2016} present in self-assembled QDs. Recent work on lattice-matched GaAs QDs has overcome this limitation, leading to high-fidelity NMR control \cite{chekhovich_nuclear_2020}, nuclear hyper-polarisation \cite{millington-hotze_approaching_2024}, enhanced electron spin dephasing times~\cite{nguyen_enhanced_2023} and spin coherence exceeding \SI{100}{\micro s} under dynamical decoupling~\cite{zaporski_ideal_2023}. 
The final requirement of a nuclear quantum register is the union of a controllable electron spin with an engineered nuclear ensemble.

In this article, we demonstrate reversible quantum state transfer between an electron spin qubit and a collective excitation of $13000$ nuclear spins in a GaAs QD. To achieve this, we introduce a way to engineer a collective nuclear state by polarizing the $^{69}$Ga and $^{71}$Ga isotopes in opposite directions. The result is that one isotope ($^{71}$Ga) is prepared in our register ground state consisting of a coherent nuclear dark state with 60\% polarisation. 
Our controlled electro-nuclear SWAP gates enable arbitrary state transfer from the electronic spin qubit to the single nuclear-magnon states. Ramsey interferometry of the register states operates as a quantum sensor to detect directly the electronic Knight field experienced by the nuclei.
When decoupled from the Knight field, the register achieves a \storagetime storage time consistent with limits set by quadrupolar broadening, promising extension beyond 20 ms~\cite{chekhovich_nuclear_2020} with the addition of nuclear control pulses. 

%
%
\begin{figure*}[t]%
\centering
\includegraphics[width=1\textwidth]{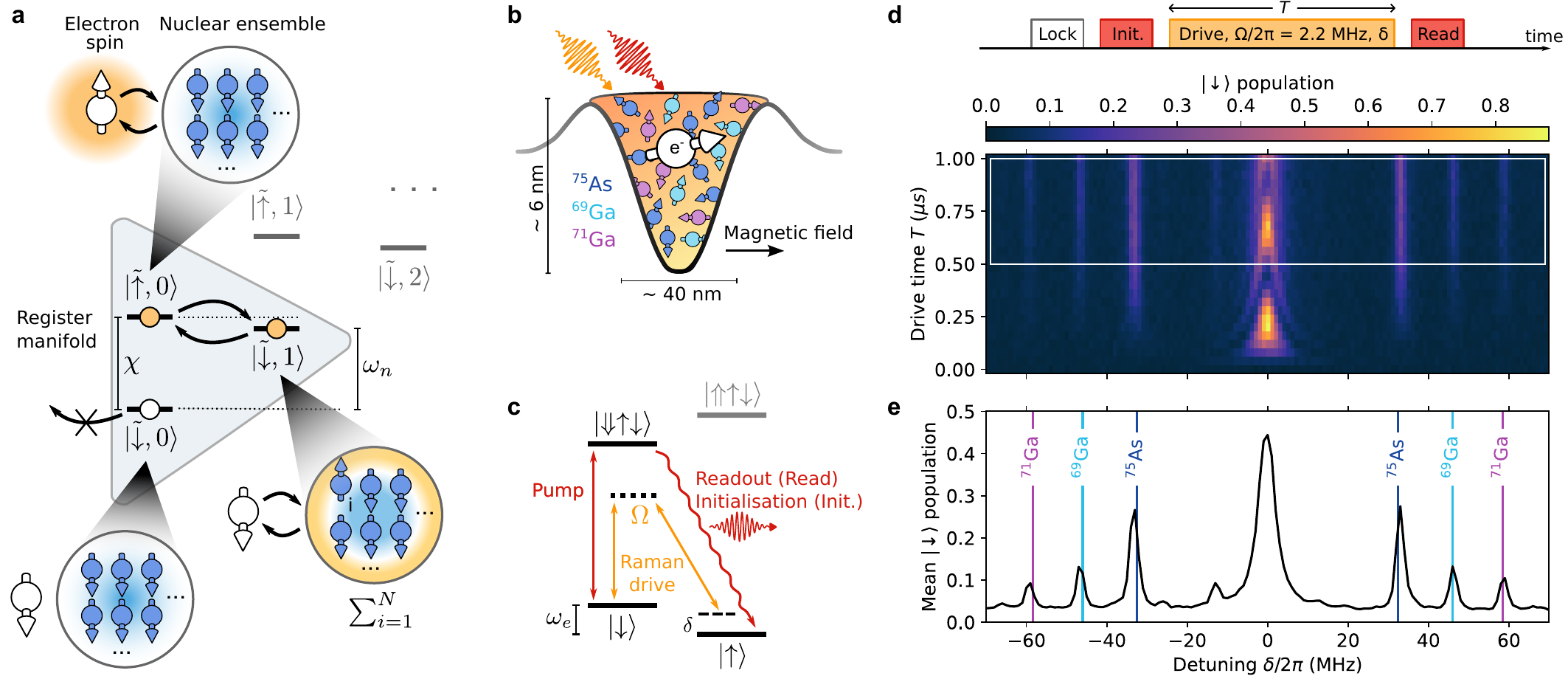}
    \caption{\textbf{Isotope-resolved electro-nuclear interface.} \textbf{a,} A central electron spin coupled to $N$ nuclear spins prepared in a dark state. The two lowest ensemble states create a register manifold where an electronic excitation (yellow) can be swapped to a collective nuclear excitation once the dressed-state splitting $\chi$ is resonant with the nuclear Larmor frequency $\omega_n$. \textbf{b,} Physical realization of a central spin system using a GaAs quantum dot hosting three nuclear species. 
    \textbf{c,} Energy level diagram containing the ground states $\ketup,\ketdown$. A pump laser (784.6 nm) resonant with the trion state $\ket{\Down\up\down}$ allows readout of $\ketdown$ and initialization into $\ketup$. The Zeeman splitting $\omega_e$ is bridged by a $\sim\!\!\SI{600}{GHz}$ detuned bi-chromatic laser (yellow arrows) with two-photon detuning $\delta$ and spin Rabi frequency $\Omega$. \textbf{d,} Measured ESR spectrum, where the electron is initialized in $\ketup$ and driven with detuning $\delta$ for drive time $T$, following nuclear polarisation locking at $I_z\sim 0$ (see sequence inset). \textbf{e,} Average signal within the white box in d. Three pairs of sidebands occur at the expected nuclear Larmor frequencies of the host material (colored lines). Other features stem from second-order processes (Supplementary Information).}
    \label{fig:1}
\end{figure*}
%
%
%

%
%
\begin{figure*}[t]%
\centering
\includegraphics[width=1\textwidth]{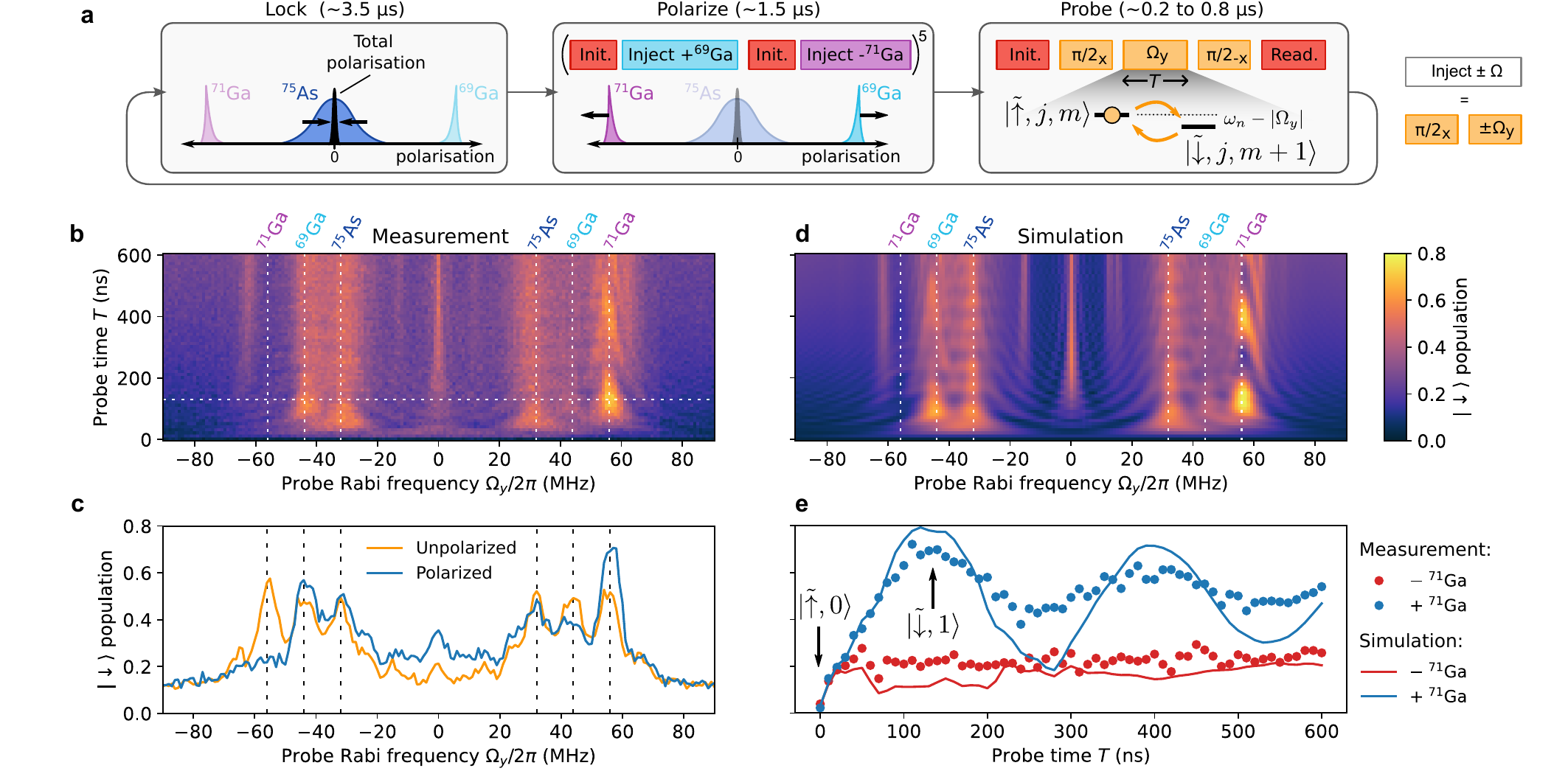}
    \caption{\textbf{Engineering a many-body dark state.} 
    \textbf{a,} Experimental pulse sequence and its effects on nuclear polarisation. 
    The polarize and probe steps use a NOVEL drive (rightmost inset) to inject polarisation into the species with $\omega_n=|\Omega_y|$. 
    In the probe step (illustrated for $\Omega_y>0$), NOVEL driving couples $\ket*{\tilde{\up},j,m}\leftrightarrow\ket*{\tilde{\down},j,m+1}$, and the final $\pi/2$ pulse maps $\ket*{\tilde{\down}}\rightarrow\ketdown$ before readout.
    \textbf{b,} Probe spectrum of the polarized nuclear ensemble as a function of probe time $T$ and Rabi frequency $\Omega_y$. 
    \textbf{c,} Probe spectra from an unpolarized (orange curve) and polarized nuclear ensemble (blue curve) after a $T=\SI{130}{ns}$ probe. The polarized trace corresponds to the dotted horizontal line in \textbf{b}. 
    \textbf{d,} Monte Carlo simulation of the probe step for an initial polarized nuclear state with $^{71}$Ga in a dark state. 
    \textbf{e,} Measured and simulated probe signals (see legend) as a function of probe time for a polarized nuclear ensemble when driving the \plusminusseventyone sidebands. Circles and lines correspond to the $^{71}$Ga vertical dashed lines in \textbf{b,d}. 
    }
    \label{fig:2}
\end{figure*}
\section{Isotope-resolved electro-nuclear interface}
Our QD system can be approximated as a central electron spin $\hat{\mathbf{S}}$ coupled to an ensemble of $N$ identical nuclear spins $\hat{\mathbf{I}}_i$. We define the collective nuclear spin operator $\hat{\mathbf{I}}=\sum_i\hat{\mathbf{I}}_i$ and work in the collective basis $\ket{j,m}$: $j\leq j_0$ is the total angular momentum where $j_0=NI$ is the maximal spin length set by the nuclear spin magnitude $I$, and $m=-j,\dots,j$ is the spin projection along $z$. For an external magnetic field along $z$, the system evolves under the approximate Hamiltonian
\begin{align}
\hat{H}=\omega_n \hat{I}_z + a_\parallel\hat{S}_z\hat{I}_z + \frac{a_\perp}{2}\hat{S}_z (\hat{I}_+ +\hat{I}_-)+\hat{H}_\text{drive}, \label{eq:Hamiltonian}
\end{align}
where $\omega_n$ is the nuclear Larmor frequency and $a_\parallel(a_\perp)$ is the collinear(non-collinear) hyperfine coupling constant. We provide the energy required for a nuclear spin flip through the rotating frame electron drive $\hat{H}_\text{drive}=\delta\hat{S}_z+\Omega\hat{S}_x$ which creates dressed electronic states $\{\ket*{\tilde{\up}}$,$\ket*{\tilde{\down}}\}$ split by $\chi=\sqrt{\delta^2+\Omega^2}$ (Fig. 1a). Setting $\chi=\omega_n$ satisfies the Hartmann–Hahn resonance condition \cite{hartmann_nuclear_1962} and leads to evolution dominated by
\begin{align}
    \tilde{H}=\frac{a_\perp\Omega}{2\chi}\left(\ketbra*{\tilde{\up}}{\tilde{\down}}\hat{I}_- + \ketbra*{\tilde{\down}}{\tilde{\up}}\hat{I}_+\right),\label{eq:main:Htilde}
\end{align}
where dressed-state spin flips inject collective nuclear excitations.
We now imagine an ensemble prepared in a coherent dark state $\ket{j,-j}$ with $j<j_0$. Further reduction of $m=-j$ is forbidden under the system symmetries~\cite{taylor_controlling_2003,gangloff_witnessing_2021}, allowing the two lowest ensemble states $\ket{0}=\ket{j,-j}$ and $\ket{1}=
\ket{j,-j+1}$ to form a closed manifold permitting deterministic quantum state transfer (Fig.~1a). For a general input state $\alpha\ket*{\tilde{\down}}+\beta\ket*{\tilde{\up}}$, correctly timed evolution under Eq. \ref{eq:main:Htilde} yields $(\alpha\ket*{\tilde{\down}}+\beta\ket*{\tilde{\up}})\ket{0}\rightarrow\ket*{\tilde{\down}}(\alpha\ket{0}+\beta\ket{1})$, thereby storing the input state in a superposition of the register ground state and a single collective excitation (a nuclear magnon). 

The physical realization of our central spin system is a QD device containing a GaAs droplet embedded in AlGaAs~\cite{huo_ultra-small_2013, huber_highly_2017, chekhovich_measurement_2017}, the lattice matching of which results in low strain and correspondingly low nuclear quadrupolar broadening of $10$ to $100$ kHz~\cite{chekhovich_nuclear_2020,zaporski_ideal_2023}. Our spin qubit is a conduction band electron confined to a volume containing $\sim 10^5$ spin-3/2 nuclei (Fig. 1b) distributed across the species $^{75}$As, $^{69}$Ga and $^{71}$Ga with abundances of 100\%, 60.1\% and 39.9\%, respectively~\cite{berglund_isotopic_2011}.
We operate the QD device at 4 K with a 4.5-T in-plane magnetic field tilted 45$\degree$ from the crystallographic axes, yielding $\omega_e/2\pi=\SI{2.5}{GHz}$ electron Zeeman splitting. In GaAs QDs, this field orientation together with the anisotropy of the electron g-factor results in the non-collinear coupling $a_\perp$~\cite{botzem_quadrupolar_2016, shofer_tuning_2024}. 
Electron spin initialization and readout are realized through resonant optical excitation, while coherent spin control is provided by a Raman scheme~\cite{gangloff_quantum_2019, bodey_optical_2019} where a two-photon detuning determines $\delta$, and $\Omega$ is controlled by power of the two-color Raman laser (Fig.~1c). 

The transformative nature of nuclear homogeneity in GaAs QDs is revealed directly through the electron spin resonance (ESR) spectrum. 
In the un-driven system at thermal equilibrium, the fluctuating nuclear polarisation $\expval{(I_z)^2}=NI(I+1)/3$ couples to the electron via the second term in Eq.~\ref{eq:Hamiltonian} resulting in a spin dephasing time $T_{2,e}^*\approx \SI{2.5}{ns}$ \cite{zaporski_ideal_2023} and an inhomogeneous ESR linewidth of 210 MHz. We overcome this limitation by preceding ESR measurements with quantum-algorithmic feedback (Fig. 1d pulse sequence) that locks $I_z$~\cite{jackson_optimal_2022,nguyen_enhanced_2023}, prolonging the electron $T_{2,e}^*$ to 290 ns and yielding a coherence-limited linewidth of 1.8-MHz FWHM (Supplementary Information).
To avoid coherent broadening, we drive the electron in the detuned regime $\delta\gg\Omega$ where the Hartmann–Hahn condition manifests as sidebands in the ESR spectrum at $\delta=\pm\omega_n$ ($\pm$ sideband) for each of the three nuclear species~\cite{gangloff_quantum_2019}. 
Figure 1d presents the measured ESR spectrum as a function of drive time. It contains electron Rabi oscillations at a rate of $\Omega/2\pi=\SI{2.2}{MHz}$ for $\delta=0$ and three pairs of symmetric sidebands at the expected nuclear Larmor frequencies. In stark contrast to previous InGaAs QDs \cite{gangloff_quantum_2019,jackson_quantum_2021,gangloff_witnessing_2021}, the individual atomic species as well as their isotopes are distinctly resolved. The relative peak amplitudes (Fig.~1e) are consistent with the species abundances and a $1/\omega_n$-rolloff associated with detuned driving. This sideband-resolved regime of a qubit-ensemble interaction makes it possible to pump the ensemble to its ground state as done in optomechanical \cite{schliesser_resolved-sideband_2009} and trapped atomic \cite{wineland_laser_1979} systems.
\section{Nuclear Dark-State Engineering}
The species resolvability allows us to maintain a locked species-summed polarization ($I_z=0$) by only actuating $^{75}$As while simultaneously engineering an undisturbed, pure nuclear state of a gallium isotope. We employ a cyclic pulse sequence (Fig.~2a) where the gallium isotopes are subjected to fast, directional polarisation injection via the NOVEL driving scheme~\cite{henstra_nuclear_1987, bodey_optical_2019}: A $\pi/2$ $\hat{\sigma}_x$ rotation followed by a $\hat{\sigma}_y$ drive with Rabi frequency $\Omega_y$ configures the electron for spin-locking. The Hartmann–Hahn condition is fulfilled for $\chi=|\Omega_y|=\omega_n$, and the drive phase $(\pm \hat{\sigma}_y)$ dictates the direction of nuclear polarisation. By alternating between periods of NOVEL drive and optical repumping of the electron spin, polarization is repeatedly transferred from the electron to the gallium isotopes. 
We choose to pump the $^{69}$Ga and $^{71}$Ga ensembles in opposite directions such that their hyperfine shifts of the ESR frequency roughly cancel. We expect this anti-polarised configuration to be highly stable, as the $^{75}$As-based feedback only needs to correct minor fluctuations around $I_z=0$.

The resulting nuclear state is probed using another NOVEL drive (Fig. 2a). This time, an additional $\pi/2$-pulse maps the dressed states back to the readout basis via $\ket*{\tilde{\downarrow}}\rightarrow \ketdown$.  
Figure 2b shows the measured spectrum as a function of probe time $T$, with Fig. 3c displaying a linecut of the spectrum at $T=\SI{130}{ns}$. It contains the same nuclear resonances as Fig. 1d but differs in several important aspects. First, probing with $\Omega_y=0$ equates to Ramsey interferometry and a slow relaxation due to $T_{2,e}^*$.
Second, the nuclear sidebands are coherently broadened by the NOVEL drive.
Third, while the sidebands of the feedback species $^{75}$As remain symmetric, the $^{69}$Ga and $^{71}$Ga sidebands are strongly asymmetric and in opposite directions as expected from an anti-polarized state. We observe a near-perfect suppression of the \minusseventyone sideband - the clearest indication to date of a nuclear dark state~\cite{gangloff_witnessing_2021}. Figure 2d shows a Monte Carlo simulation (Supplementary Information) where the initial nuclear states are sampled from a thermal, near-dark, and fully dark state for $^{75}$As, $^{69}$Ga and $^{71}$Ga, respectively, showing remarkable agreement with the measured NOVEL spectrum. 

The observation of a dark state indicates a high level of purity of the $^{71}$Ga ensemble. The \plusseventyone~sideband corresponds to driving the $\ket*{\tilde{\up},0}\leftrightarrow\ket*{\tilde{\down},1}$ transition (c.f. Fig.~1a) and yields clear Rabi oscillations (Fig. 2e), further signifying the nuclear state purity and the coherence of the electro-nuclear coupling. The peak electron spin inversion at 130 ns corresponds to the injection of a single magnon at a Rabi frequency $\Omega_\text{mag}/2\pi=\SI{3.8}{MHz}$. For the nuclear dark state, the theoretical magnon injection rate is $\Omega_\text{mag}=a_\perp\sqrt{j/2}$ (Supplementary Information). Measuring $a_\perp^{(71)}/2\pi=\SI{50}{kHz}$ from an unpolarized NOVEL spectrum allows us to extract the dark state spin length $j=0.6\times j_0$, where $j_0$ is the independently measured maximum spin length (Supplementary Information). Notably, $j/j_0$ is $100$-fold larger than the thermal expectation value $1/\sqrt{N}$ \cite{wesenberg_mixed_2002}, implying that the polarisation step pumps the total angular momentum. The initialization of a nuclear dark state $j<j_0$ also implies considerable entanglement within the nuclear many-body system \cite{gangloff_witnessing_2021} and signifies that magnons are injected much faster than the nuclear decorrelation time.

The magnon Rabi oscillations in Fig.~2e are damped by two predominant mechanisms. First, the spectral overlap of neighboring nuclear sidebands (evident from Fig. 2b) results in dephasing of the \plusseventyone transition. Second, spin relaxation proportional to laser power~\cite{bodey_optical_2019, appel_entangling_2022} (visible as an increasing background at high $|\Omega_y|$ in Fig. 2b) further damps the \plusseventyone oscillation. These two error mechanisms additionally imply a $\sim\!\!2\%$ electron $\pi$-pulse error and unwanted electron inversion when driving the suppressed \minusseventyone transition (red circles in Fig.~2e), with both effects being of consequence to state transfer.
A simulation including these errors together with a 0.9\% spin initialization error produces the solid blue curve in Fig. 2e.
The remaining discrepancy between the measured and simulated \plusseventyone oscillations in Fig.~2e may be due to additional control pulse errors or inhomogeneity of the dark state spin length $j$. Nonetheless, the achieved visibility of the magnon Rabi oscillations is already sufficient to demonstrate quantum state transfer.

%
%
\begin{figure}[]%
\centering
\includegraphics[width=1\columnwidth]{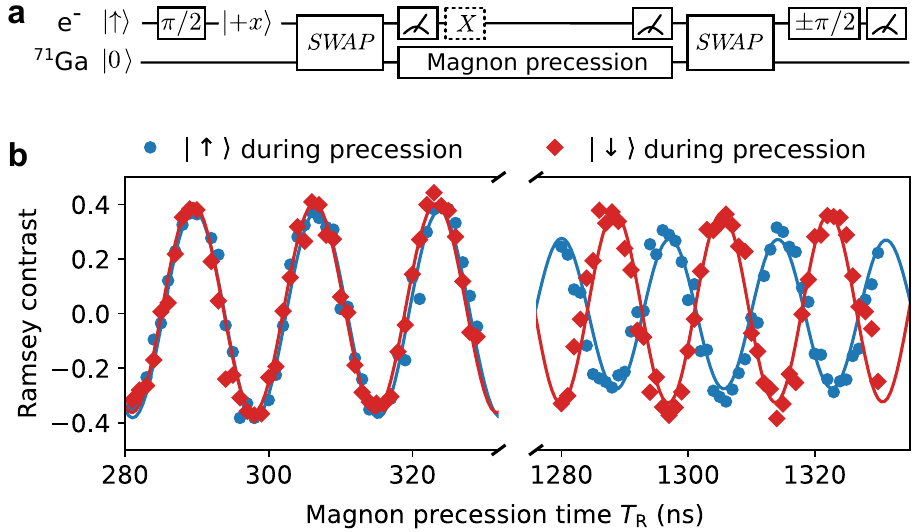}
    \caption{\textbf{Sensing the Knight field with a single magnon.} \textbf{a,} Quantum circuit for magnon Ramsey interferometry. The protocol is preceded by the lock and polarize steps in Fig.~2a. 
    \textbf{b,} Magnon Ramsey contrast $c=(p_+-p_-)/(p_++p_-)$, where $p_\pm$ is the probability of projecting the electron onto $\ket{\pm x}$ during the final readout in \textbf{a}. 
    The two datasets (blue circles and red diamonds) denote different states of the electron spin during magnon precession. For both datasets, a damped sinusoid is fit to the entire time series (left and right) to estimate the electron spin-dependent precession frequency and thus the Knight shift.
    \label{fig:3}}
\end{figure}
\section{Quantum state transfer and information storage}
We now use the 130-ns NOVEL drive resonant with $^{71}$Ga as an electro-nuclear SWAP gate. We first investigate the coherence dynamics of the nuclear states $\ket{0}$ and $\ket{1}$ through a magnon Ramsey interferometry protocol comprised of two SWAP gates with an intermediate delay, illustrated in Fig. 3a. The first SWAP gate maps the electron superposition state $\ket{+x}=(\ketdown+\ketup)/\sqrt{2}$ to the nuclear state $(\ket{0}+\ket{1})/\sqrt{2}$, i.e. a coherent superposition of the ensemble dark state and a single nuclear magnon.
This state precesses in the magnetic field at the nuclear Larmor frequency for time $T_\text{R}$ after which it is transferred back via the second SWAP gate to the electron, which is measured in the $x$-basis. An electron reset operation immediately after the first SWAP guarantees that the quantum state is stored in the $^{71}$Ga mode, and an optional $x$-gate toggles the electron state during nuclear precession. A second reset repolarizes the electron immediately before the second SWAP.
Figure 3b shows the measured Ramsey fringes and reveals a nuclear state precession frequency $\nu^\up=\SI{58.560(9)}{MHz}$ or $\nu^\down=\SI{59.060(8)}{MHz}$ when the electron is in $\ketup$ or $\ketdown$ during precession, respectively. The mean frequency $(\nu^\up+\nu^\down)/2=\SI{58.810(6)}{MHz}$ is in excellent agreement with the expected $^{71}$Ga Larmor frequency $\omega_n^{(71)}/2\pi=\SI{58.41}{MHz}$~\cite{malinowski_notch_2016} considering not more than $1\%$ magnetic field calibration error. 
The frequency difference $\delta\nu=\nu^\down-\nu^\up=\SI{0.500(12)}{MHz}$ is an explicit measurement of the hyperfine shift of nuclear energy levels caused by the electron $\{\ketup,\ketdown\}$ states. In other words, we leverage the magnon as a precise quantum sensor of the magnetic field (the Knight field~\cite{lai_knight-field-enabled_2006,sallen_nuclear_2014}) due to the electron spin. 
This measurement directly yields the effective number of $^{71}$Ga nuclei as $N^{(71)}=1.54\times A^{(71)}c^{(71)}/(2\pi\delta\nu)=1.3\cdot 10^4$ where 1.54 is a geometrical factor compensating the non-uniform nuclear coupling (Supplementary Information) and $A^{(71)}$ and $c^{(71)}$ are the material hyperfine constant and abundance of $^{71}$Ga, respectively. 

Having demonstrated that quantum coherence can be transferred to the nuclei and back, we next focus on its applicability to quantum information storage. Figure 4a displays a pulse protocol for full quantum process tomography using $\{\ket{\pm x},\ket{\pm y},\ket{\pm z}\}$ as electron qubit input states and bases for projective readout. We operate at a storage time of 290 ns (first Ramsey peak in Fig. 3b) corresponding to an integer number of nuclear precessions. Figure 4b shows the detection probability for all 36 input and readout combinations for the ideal, measured, and simulated cases. In the ideal case, all processes on the diagonal succeed with unity probability. The measurements yield average raw contrasts of $C_x=0.348(15), C_y=0.344(16), C_z=0.423(15)$ and a total fidelity \cite{dutt_quantum_2007} of $F=(1+\expval{C})/2=0.686(4)$, where $\expval{C}=(C_x+C_y+C_z)/3$. This total fidelity encompassing all errors from two SWAP gates, spin initialization and single qubit rotations still exceeds the classical limit of $2/3$~\cite{dutt_quantum_2007}. The storage fidelity of $\ket{-z}$ notably exceeds that of $\ket{+z}$ as the latter state involves injecting and retrieving a magnon while the former does not. Simulating the full protocol with our previous Monte Carlo approach qualitatively reproduces the experimental results including the $\ket{-z}$ bias and yields a marginally higher fidelity $F=0.730$. This indicates that the error mechanisms previously identified account for the measured infidelity, offering guidance toward improvement.
%
%
\begin{figure}[h!]%
\centering
\includegraphics[width=1\columnwidth]{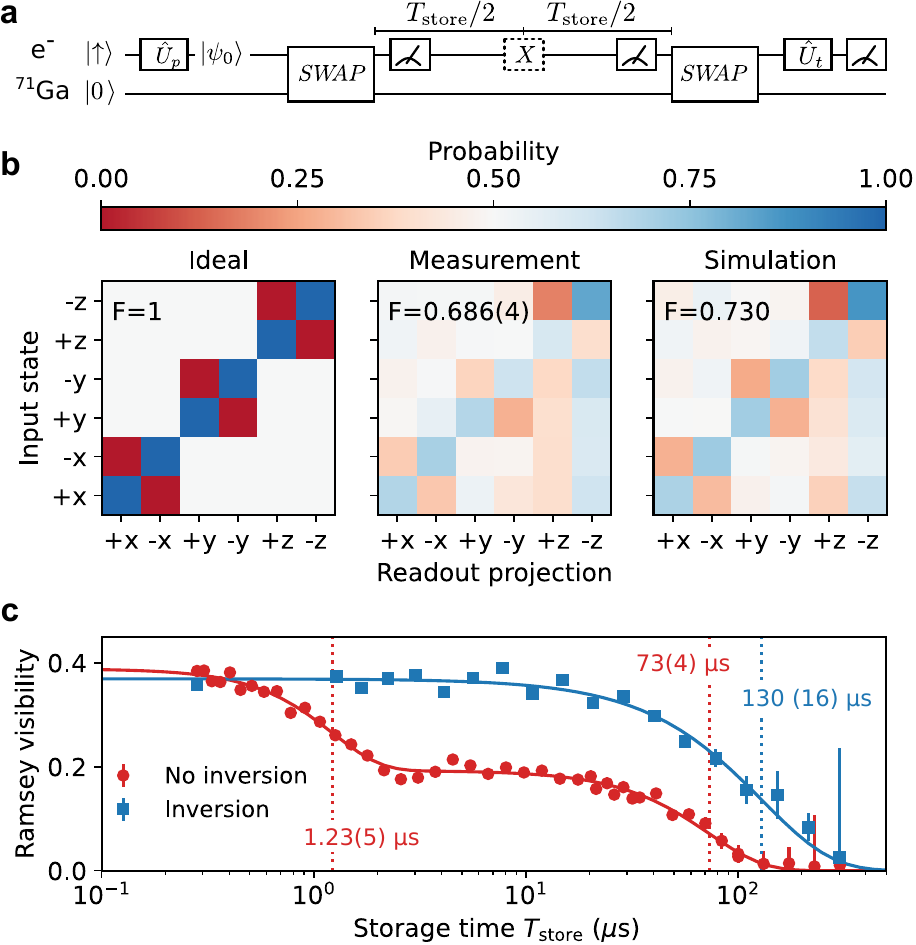}
    \caption{\textbf{State-transfer tomography and information storage.} 
    \textbf{a,} Quantum circuit for storage and retrieval of electron spin states in a nuclear register. An electron spin rotation $\hat{U}_p$ prepares an input state $\ket{\psi_0}$ which is transferred to the register. After a storage time $T_\text{store}$, the state is retrieved, and the electron is projected onto a single axis determined by $\hat{U}_t$. An optional electron inversion (dotted box) cancels the Knight shift during storage. The protocol is preceded by the lock and polarize steps in Fig. 2a.
    \textbf{b,}
    Quantum state tomography of the final electron state as a function of input and readout projections. The three panels represent the results of an ideal memory, the raw measurement, and a Monte Carlo simulation.
    \textbf{c,} Measured Ramsey fringe visibility over two Larmor periods when storing $\ket{\psi_0}=\ket{x}$. Red circles are fit with $V(t)=Ae^{-(t/T_a)^\alpha}+Be^{-(t/T_b)^\beta}$ yielding $\{A,T_a,\alpha,B,T_b,\beta\}=\{0.196(11),\SI{1.23(5)}{\micro s},1.9(3),0.193(4),\SI{73(5)}{\micro s},1.7(2)\}$. Including the electron inversion yields the blue squares which are fit with $V(t)=Ae^{-(t/T_a)^\alpha}$ yielding $\{A,T_a,\alpha\}=\{0.369(6),\SI{130(16)}{\micro s},1.3(2)\}$. $T_a$ and $T_b$ are indicated by the dotted lines.
    \label{fig:4}}
\end{figure}

Finally, having demonstrated arbitrary state transfer between the electron spin qubit and the collective-state nuclear register, we benchmark the storage time offered by the register.
Figure 4c is the magnon Ramsey visibility for qubit state $\ket{\pm x}$ as a function of storage time $T_\text{store}$ (see Supplementary Information for $\ket{z}$ storage). 
The red curve, corresponding to free evolution for time $T_\text{store}$, reveals a clear two-stage dephasing with characteristic times of \SI{1.23(5)}{\micro s} and \SI{73(4)}{\micro s}.  
Based on our measurement of the Knight field, we expect a dephasing of the magnon on a timescale $\sim a_\parallel^{-1}$ due to the spatial gradient of $a_\parallel$ across the nuclear ensemble owing to the quasi-Gaussian electron wavefunction within the QD. This timescale is fully commensurate with the short time scale observed. As the Knight field is static, its effect is averaged to zero by inverting the electron state halfway through the protocol (dotted gate in Fig.~4a). With this inversion pulse, the blue squares in Fig.~4c indeed reveal a disappearance of the short-term dephasing, and the available storage time extends to $T_{2,n}^*=\storagetime$ without significant reduction of the initial visibility. NMR measurements of similar QDs report a $\SI{7}{kHz}$ FWHM of the $^{69}$Ga satellite transitions~\cite{millington-hotze_approaching_2024} corresponding to $T_{2,n}^*=\SI{120}{\micro s}$ for $^{71}$Ga, suggesting that our $T_{2,n}^*$ is likewise limited by quadrupolar broadening. 

To further improve the storage time, NMR control can transfer quantum information to the narrow $\ket{-1/2}\leftrightarrow\ket{+1/2}$ nuclear transition and perform dynamical decoupling, which has already resulted in 20-ms coherence in GaAs QDs~\cite{chekhovich_nuclear_2020}. Further nuclear environmental control measures including repeated electron inversion or fast charge control may protect against higher-order electron-mediated dephasing~\cite{wust_role_2016}. Regarding the full process fidelity, the spectral overlap of different species under NOVEL drive is the main imperfection, currently leading to a 23\% simulated infidelity. This error can be suppressed by reducing $a_\perp$ through magnetic field alignment \cite{botzem_quadrupolar_2016} or net nuclear polarisation~\cite{shofer_tuning_2024}. Eliminating $^{69}$Ga through isotopic purification will further reduce overlap. In the ideal case of a $^{71}$Ga/$^{75}$As QD with both species in $j=0.6\times j_0$ dark states, the simulated overlap error only induces a$1.7\%$ infidelity for the current value of $a_\perp$. Alternatively, Hamiltonian engineering may be used to achieve species-selective transfer while remaining insensitive to electron $T_{2,e}^*$~\cite{denning_collective_2019}. The laser-induced electron spin relaxation~\cite{bodey_optical_2019, appel_entangling_2022} currently contributes an 8.5\% infidelity which can likely be improved through device design and enhanced optical mode matching to reduce the optical power needed for qubit control. 
\section{Conclusion and outlook}    
In summary, we have demonstrated a functional quantum register based on a nuclear many-body system interfaced with an electron spin qubit. When operated as a memory, this many-body register transforms QDs into fully fledged quantum network nodes. The current storage time of $T_{2,n}^*=\storagetime$ is already sufficient for fast protocols such as 2d-cluster state generation~\cite{buterakos_deterministic_2017, michaels_multidimensional_2021} and Bell state analysers~\cite{witthaut_photon_2012}. 
$T_{2,n}^*$ already greatly exceeds the $T_{2,e}^*$ of the electron spin qubit, and while it is equivalent to the dynamically decoupled electron spin coherence time~\cite{zaporski_ideal_2023}, NMR control will extend the nuclear coherence time to the 20 ms regime~\cite{chekhovich_nuclear_2020}.
Another opportunity is presented by the remaining nuclear species, which can operate in parallel to increase the quantum information storage capacity of our device.
Beyond quantum node development, our demonstrated control of a central spin system in the coherent regime enables foundational studies of collective phenomena including super-radiant nuclear spin dynamics, time-crystalline behavior~\cite{frantzeskakis_time-crystalline_2023} and engineering of many-body singlets \cite{zaporski_many-body_2023}.

\section*{Acknowledgements}
We would like to thank E. Chekhovich for fruitful discussions, as well as M. Tribble for device fabrication advice. We additionally acknowledge support from the US Office of Naval Research Global (N62909-19-1-2115;M.A.), the EU Horizon 2020 FET Open project QLUSTER (862035; M.A. and C.L.G), the EU Horizon 2020 research and innovation program under Marie Sklodowska-Curie grant QUDOT-TECH (861097; M.A.), the Royal Society (EA/181068; C.L.G), Qurope (899814; A.R.), ASCENT+ (871130; A.R.), the Austrian Science Fund (FWF; 10.55776/COE1; A.R.), the EU NextGenerationEU (10.55776/FG5; A.R.), and SERB India (CRG/2023/007444; S.M.). L.Z. acknowledges support from the EPSRC DTP (EP/R513180/1) and A.G. from a Harding scholarship and a Christ's College scholarship. D.A.G acknowledges a Royal Society University Research Fellowship. C.L.G. acknowledges a Dorothy Hodgkin Royal Society Fellowship. \par

\bibliography{MagnonPaperBibliography_fixed.bbl}
\newpage

\clearpage

\end{document}


\newcommand{\TitleName}{Supplementary information: Many-body quantum register for a spin qubit}
\title{\TitleName}

\newcommand{\AffCam}{Cavendish Laboratory, University of Cambridge, J.J. Thomson Avenue, Cambridge, CB3 0HE, UK}
\newcommand{\AffLinz}{Institute of Semiconductor and Solid State Physics, Johannes Kepler University, Altenberger Str. 69, Linz 4040, Austria}

\author{Martin Hayhurst Appel}
\affiliation{\AffCam}
\author{Alexander Ghorbal}
\affiliation{\AffCam}
\author{Noah Shofer}
\affiliation{\AffCam}
\author{Leon Zaporski}
\affiliation{\AffCam}

\author{Santanu Manna}
\affiliation{\AffLinz}
\author{Saimon Filipe Covre da Silva}
\affiliation{\AffLinz}

\author{Urs Haeusler}
\affiliation{\AffCam}
\author{Claire Le Gall}
\affiliation{\AffCam}
\author{Armando Rastelli}
\affiliation{\AffLinz}
\author{Dorian A. Gangloff}
\email[Correspondence to: ]{dag50@cam.ac.uk}
\affiliation{\AffCam}
\author{Mete Atat\"ure}
\email[Correspondence to: ]{ma424@cam.ac.uk}
\affiliation{\AffCam}

\date{\today}

\maketitle 

\tableofcontents
\newpage
\section{Experimental setup}
\subsection{Sample details}
\begin{figure*}[h]%
\centering
\includegraphics[width=1\textwidth]{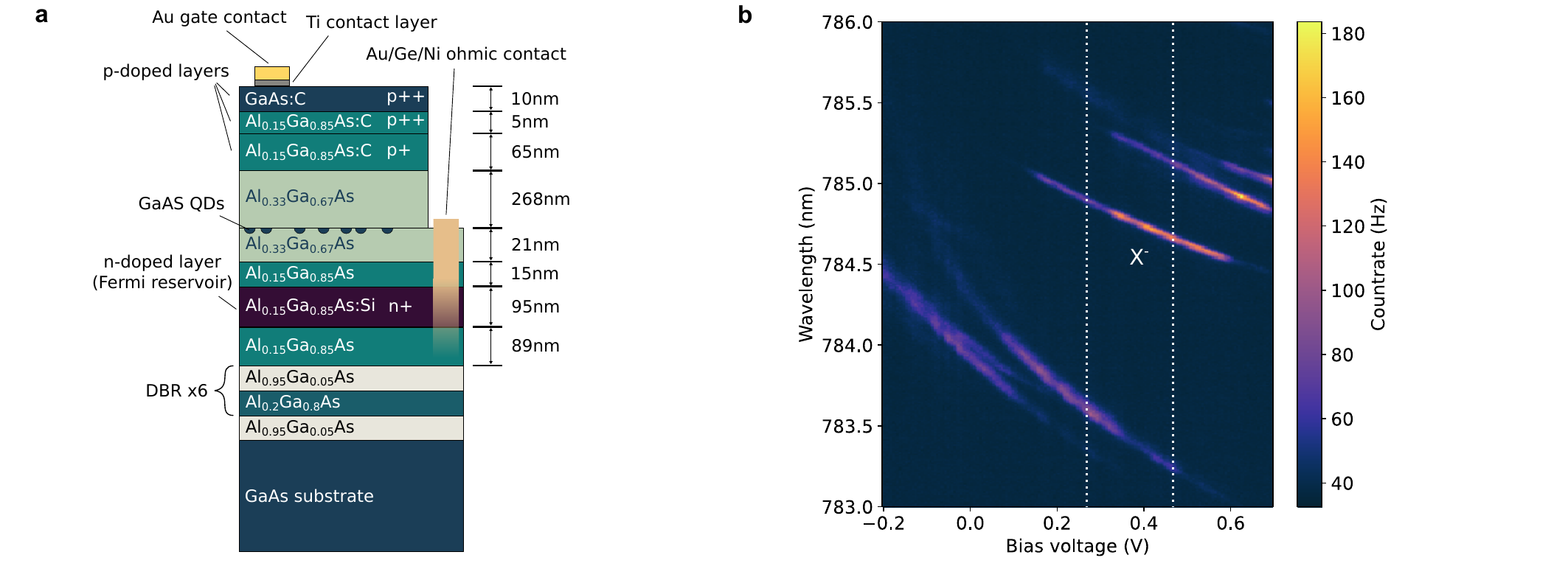}
    \caption{\textbf{a,} Schematic of the QD heterostructure. \textbf{b,} Photoluminescence spectrum of the studied QD. The QD is excited above bandgap with a 638 nm laser and the emission is resolved on a spectrometer. The emission line centered at 784.6 nm originates from the negative trion X$^{-}$ of a singly negatively charged QD. This charge state is stable under resonant excitation within the voltage range indicated by the dotted lines.}
    \label{fig:device}
\end{figure*}

Our quantum dot (QD) structure is grown via molecular beam epitaxy using Al-droplet etching and GaAs infilling to define the QDs. We refer to Ref. \cite{da_silva_gaas_2021} for growth details. The QD device consists of a p-i-n diode with the GaAs QDs embedded in the intrinsic layer. The diode heterostructure is presented in Fig. \ref{fig:device}a and is identical to the one utilized in Ref. \cite{zaporski_ideal_2023} except for the thickness of the AlGaAs barrier immediately below the QD layer which was increased from 15 nm to 21 nm. To increase the light-matter coupling, the heterostructure contains a DBR mirror (6 pairs), and a super-hemispherical zirconia solid immersion lens is attached to the top of the heterostructure.
By applying a weak forward bias across the diode, we can deterministically charge the QD.  Fig. \ref{fig:device}b shows a voltage-dependent photoluminescence measurement revealing discrete charge states including the negative charge state utilized in this work. 
\clearpage
\subsection{Strain estimation}\label{sec:strainEstimation}
\begin{figure*}[h!]%
\centering
\includegraphics[width=0.66\textwidth]{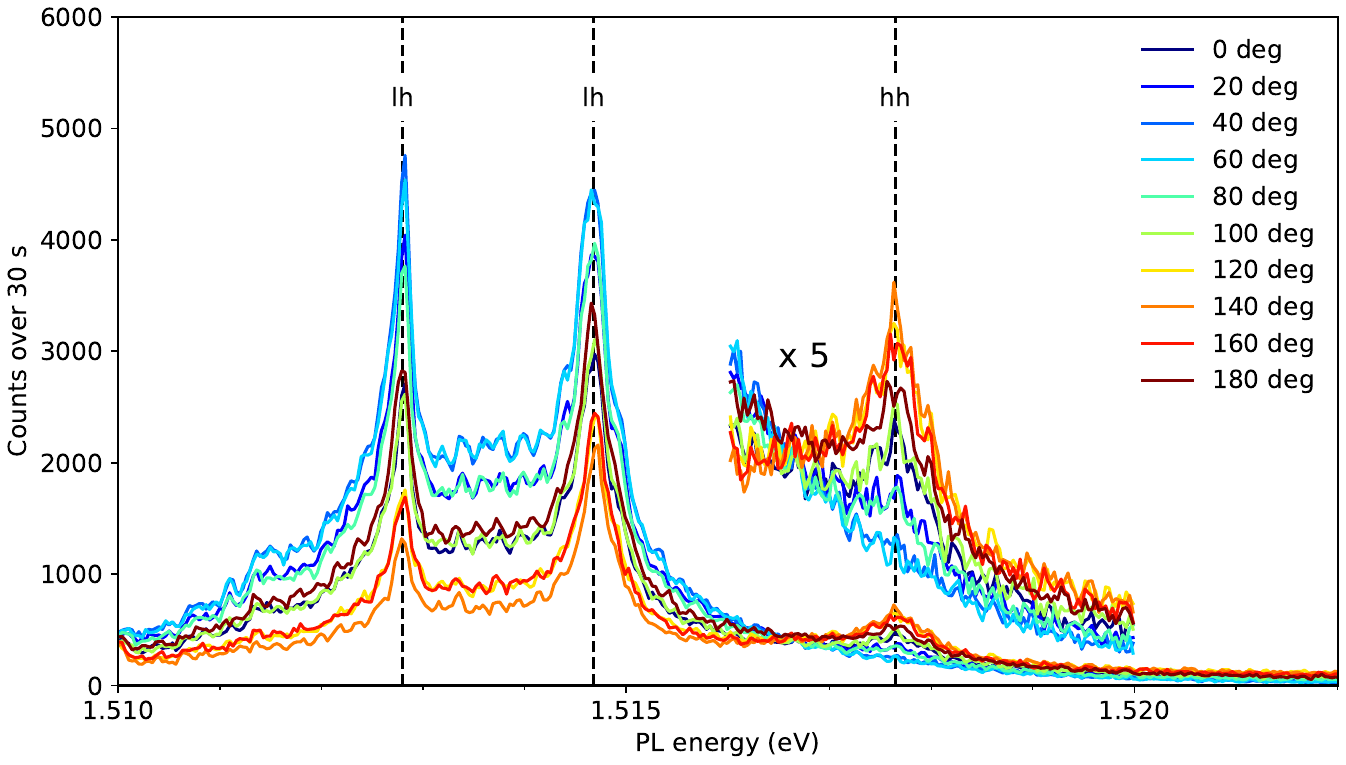}
    \caption{Photoluminescence measurement of the free GaAs exciton. The sample area containing the measured QD is excited with a 638 nm laser and the emission is resolved on a spectrometer. The legend indicates the orientation of the linear collection axis. For clarity, the spectra are also plotted with 5$\times$ magnification in the vicinity of the hh peak. }
    \label{fig:FreeExcitionPL}
\end{figure*}

We here briefly evidence that the non-collinear electro-nuclear coupling utilized in this work does not originate from strain.
We follow Ref. \cite{chekhovich_cross_2018} and use the splitting between the light hole (lh) and heavy hole (hh) emission lines of the free GaAs exciton as a proxy for strain. Fig. \ref{fig:FreeExcitionPL} shows photoluminescence measurements in which we observe a weak, linearly polarized hh emission and a doublet of partially polarized peaks corresponding to lh emission. We expect this emission to originate from the GaAs substrate in our device (bottom layer in Fig. \ref{fig:device}a). Based on the low-energy lh peak, we extract a maximal lh-hh splitting of 4.9 meV. According to Ref. \cite{chekhovich_cross_2018}, this corresponds to a nuclear quadropolar shift $B_Q/2\pi$ of 89 kHz. Based on the single nuclear hyperfine coupling $a/2\pi=\SI{0.342}{MHz}$ and Larmor frequency $\omega_n/2\pi=\SI{58.41}{MHz}$ of the $^{71}$Ga nuclear storage mode at 4.5 T (see table \ref{tab:nuclearconstants}), strain would lead to a non-collinear coupling \cite{gangloff_quantum_2019} of $a_\perp/2\pi=aB_Q/(2\pi\omega_n)=0.50$ kHz which is less than 1\% of the $a_\perp/2\pi=\SI{51}{kHz}$ estimated in section \ref{sec:NovelSim}. For this reason, we neglect strain effects in our modeling of the electro-nuclear interface.
\clearpage
\subsection{Optical setup}
\begin{figure*}[h!]%
\centering
\includegraphics[width=1\textwidth]{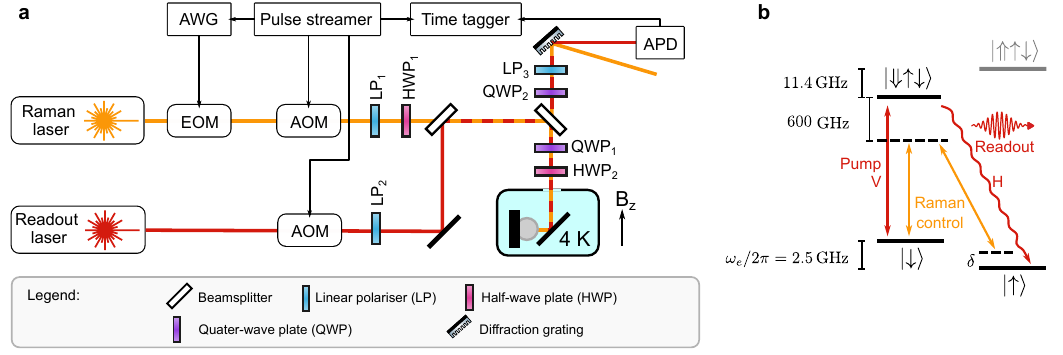}
    \caption{\textbf{a,} Optical setup. Two lasers are combined and sent onto the QD contained in a 4 K cryostat. Abbreviations are defined in the figure legend. The Raman control laser is made bichromatic by modulation with the AWG-driven EOM. The waveplates HWP$_1$ and QWP$_1$ result in a circularly polarised Raman laser field on the QD. The readout laser hits the fast axis of QWP$_1$ and retains its linear polarisation. HWP$_2$ is used to match the readout laser polarisation to the V-polarized transition (cf. panel b). The polarization optics QWP$_2$ and LP$_3$ are used to reject the resonant laser reflection. Lenses and optical fibers have been omitted from the diagram. \textbf{b,} Energy level diagram of a negatively charged QD including relevant energy splittings of the measured QD. H and V indicate the linear polarisations of the optical dipoles with V pointing along the external magnetic field.}
    \label{fig:optics}
\end{figure*}

The QD device is held at 4 K inside a helium bath cryostat where a superconducting magnet produces a 4.5 T magnetic field in the z-direction (Fig. \ref{fig:optics}a). By mounting the QD sample sideways and additionally rotating it 45$\degree$ around the optical axis, we obtain a magnetic field that is transverse to the QD growth axis and 45$\degree$ in between the crystallographic [$110$] and [$\Bar{1}10$] axes. Figure \ref{fig:optics}a illustrates the optical setup. A resonant readout laser and a Raman laser are combined on a beam splitter and sent into the cryostat. The Raman laser is detuned 600 GHz from the trion states and is circularly polarized to avoid AC-stark shifts of the electron spin during driving~\cite{press_complete_2008} and to achieve spin Rabi frequencies up to 100 MHz. The readout laser is linearly polarized such that it only drives the $\ketdown\leftrightarrow \ket{\Downarrow\up\down}$ transition and not the orthogonally polarized $\ketup\leftrightarrow\ket{\Downarrow\up\down}$ transition, see. Fig. \ref{fig:optics}b. Suppressing the latter transition significantly increases the spin initialization fidelity as this transition can otherwise off-resonantly repump the spin. 
We employ cross-polarisation to reject the resonant readout laser. Due to the linear excitation scheme, we only collect the H-polarized anti-Stokes Raman scattering. 
The polarisation-filtered emission is coupled into a fiber and spectrally filtered by a diffraction grating (30 GHz FWHM) to remove the Raman laser reflection. Finally, the QD emission is detected on an avalanche photodiode (APD) and time tagged with a Swabian Time Tagger 20. 
\\ \\
Our Raman scheme is based on the modulation of a CW-laser. The laser is modulated by a fibre-coupled amplitude electro-optical modulator (EOM) which is locked to its interferometric minimum and driven by microwave pulses generated from an arbitrary waveform generator (AWG). Driving the EOM at frequency $\omega_{\mu w}$ and phase $\varphi$ generates a pair of optical sidebands with splitting $2\omega_{\mu w}$, phase difference $2\varphi$, and resulting two-photon detuning $\delta=\omega_e-2\omega_{\mu w}$ (Fig. \ref{fig:optics}b). This control scheme is further elaborated in Ref. \cite{bodey_optical_2019}. Due to the modest $\omega_e$, we directly synthesize the microwave pulses in the time domain.

An acousto-optical modulator (AOM) placed after the EOM is used to stabilize the Raman power and to block EOM laser leakage during periods without Raman drive. An AOM is additionally used to create the resonant pumping pulses from the readout laser. The AOMs, AWG, and time tagger are all triggered by a Swabian Pulse Streamer. To ensure timing accuracy over long ($>\SI{100}{\micro s}$) histograms, the Time Tagger and Pulse Streamer are locked to the AWG's 10 MHz clock.

\clearpage
\subsection{Spin readout and initialisation}\label{sec:SpinInit}
\begin{figure*}[h!]%
\centering
\includegraphics[width=0.6\textwidth]{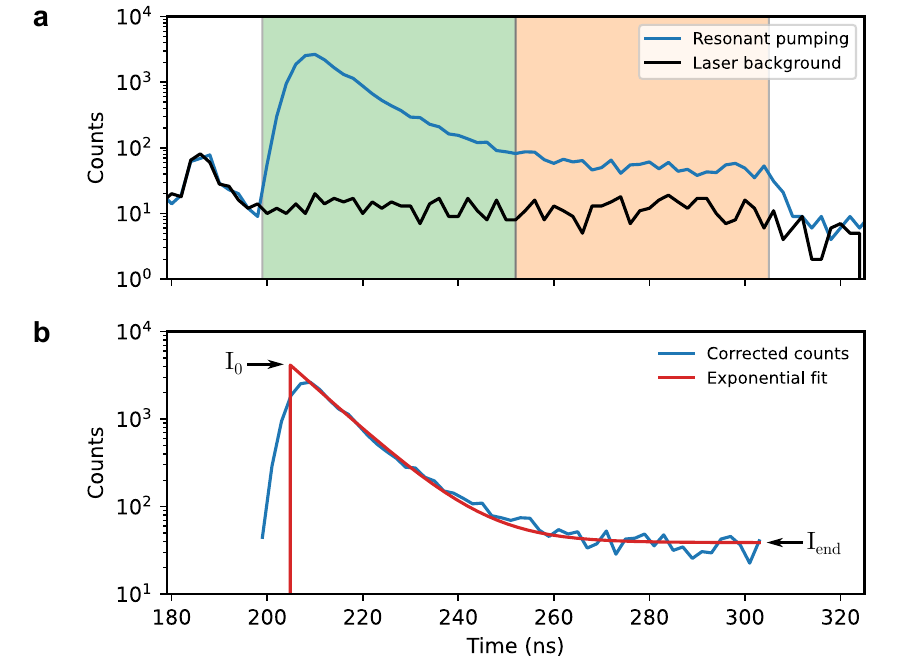}
    \caption{\textbf{a,} Histogram of a resonant readout pulse given a bright electron spin state (blue curve). The difference between the integrated counts in the green and orange areas constitutes the readout counts. The pulse at 190 ns is an optical reflection of a Raman $\pi$-pulse. By recording a histogram with a non-resonant bias voltage, we additionally estimate a background histogram (black curve). 
    \textbf{b,} Corrected histogram obtained by subtracting the histograms in \textbf{a}. The histogram is fit with the model $I(t)=\Theta(t-t_0)\left[(I_0-I_\text{end})e^{-(t-t_0)\Gamma}+ I_\text{end}\right]$, where $\Theta(t)$ is the Heaviside function and $t_0$ is chosen to ensure an equal area of the fit and the data. We further extract a spin pumping time $1/\Gamma=\SI{8.88(15)}{ns}$. }
    \label{fig:opticalPumping}
\end{figure*}

We use 100 ns long resonant pumping pulses to initialize the QD in $\ketup$ and to read out the $\ketdown$ state.
Figure \ref{fig:opticalPumping}a shows a histogram of detected QD fluorescence during a pumping pulse given an initial $\ketdown$ state. The initial rise time is a result of the $\approx 6$ ns rise time of the AOM used for pulsing. The readout counts reported in the manuscript are acquired by subtracting the counts in the second readout window from the first readout window (colored areas in Fig. \ref{fig:opticalPumping}a). This effectively subtracts the constant background owing to laser scatter and residual QD fluorescence. 

We now estimate the fidelity of spin initialization. We first estimate the background originating from laser scatter and detector dark counts by switching the QD to a non-resonant charge state. By subtracting this background, we obtain a corrected spin pumping histogram (Fig. \ref{fig:opticalPumping}b). We fit a single exponential decay to this histogram to estimate the initialization fidelity $F_\text{init}=\bra{\up}\hat{\rho}_\text{end}\ket{\up}$, where $\hat{\rho}_\text{end}$ is the state after pumping. $F_\text{init}$ can be estimated from \cite{appel_coherent_2021} 
\begin{align}
    F_\text{init} \geq 1-\frac{I_\text{end}}{I_0},
\end{align}
where $I_0$ and $I_\text{end}$ are the fluorescence intensities at the start and the end of the pumping, respectively. As this model assumes a temporally square pumping pulse, we extend the exponential fit in Fig. \ref{fig:opticalPumping}b backward in time until its area matches the measured area. From this fit, we extract $I_0=4146(72)$, $I_\text{end}=38.7(1.9)$ and $F_\text{init}\geq 99.07(5)\%$. This estimate constitutes a lower bound as we have assumed perfect initialization of $\ketdown$ prior to pumping.
\clearpage
\subsection{Pulse sequences}
\begin{figure*}[h!]%
\centering
\includegraphics[width=1\textwidth]{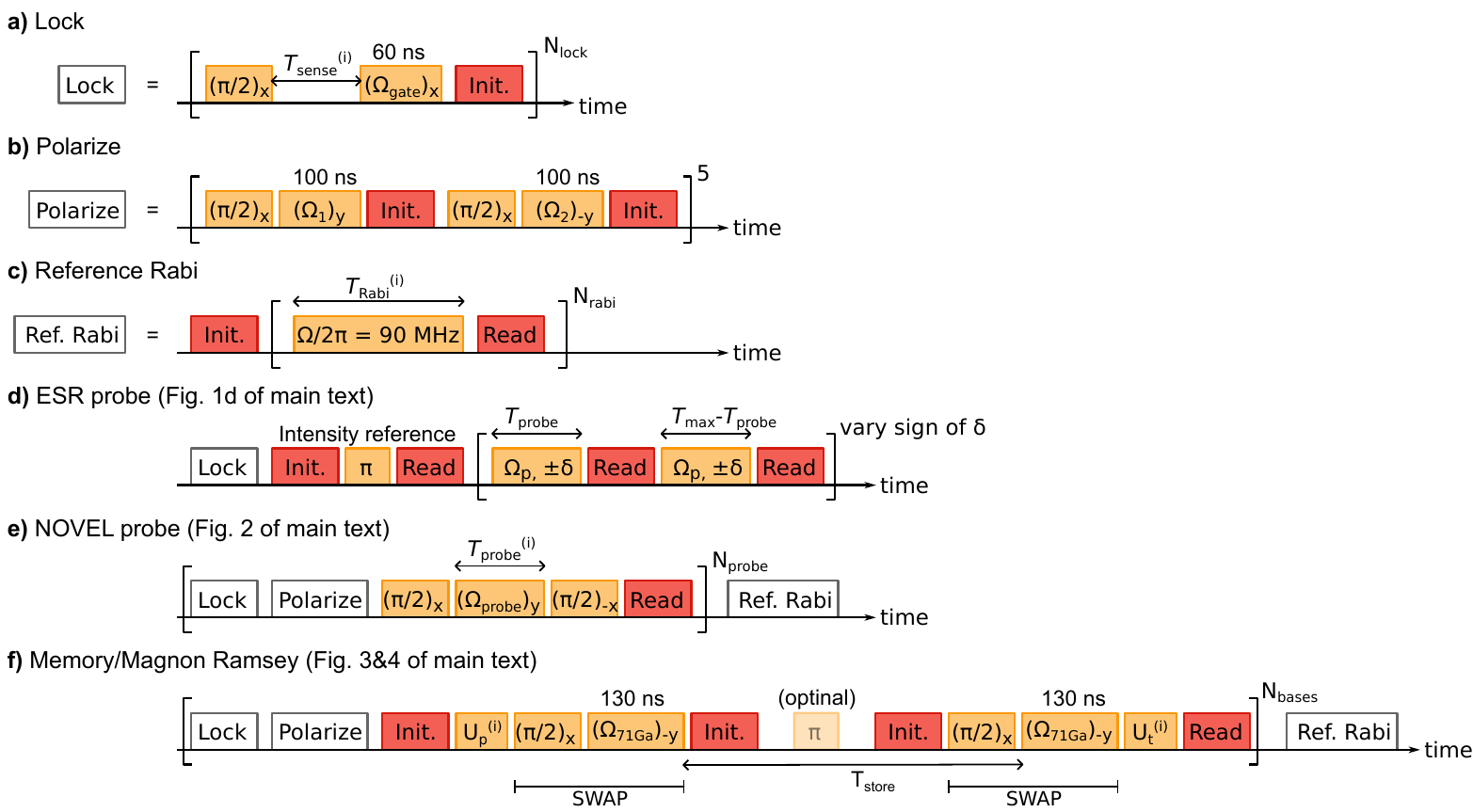}
    \caption{Experimental pulse sequences. All $\pi$ and $\pi/2$ pulses use a $\Omega/2\pi=\SI{90}{MHz}$ Rabi frequency and 5.6 ns and 2.7 ns durations, respectively. Readout (Read) and initialization (Init.) pulses are both 100 ns long. 
    Additional comments are given in the supplementary text.}
    \label{fig:pulseSequences}
\end{figure*}

Figure \ref{fig:pulseSequences} details the pulse sequences used in this work.

The algorithmic locking sequence in Fig. \ref{fig:pulseSequences}a is explained in Ref. \cite{jackson_quantum_2021}. $\Omega_\text{gate}\approx \omega_n^{^{75}As}=2\pi\times \SI{32.5}{MHz}$ is used to ensure selective activation of arsenic. For the electron spin resonance (ESR) experiment, we use $N_\text{lock}=30$ steps with a sensing time $T_\text{sense}$ linearly chirped from 30 to 185 ns. We find this saturates the electron $T_2^*$. For all subsequent measurements, we use a shorter lock step with $N_\text{lock}=12$ and $T_\text{sense}$ chirped from 60 ns to 162 ns.

For the polarize step in Fig. \ref{fig:pulseSequences}b, we choose $\Omega_1/2\pi=\SI{44}{MHz}$ and $\Omega_2/2\pi=\SI{56}{MHz}$ to target $^{69}$Ga and $^{71}$Ga, respectively. The opposite drive phases $+y$ and $-y$ antipolarize the two species as explained in the main text. 

\begin{figure*}[h]%
\centering
\includegraphics[width=0.6\textwidth]{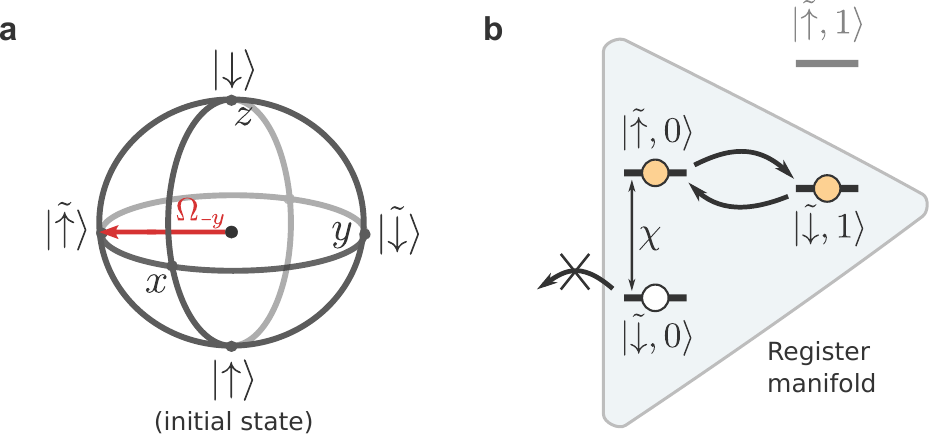}
    \caption{Electron spin rotations used for state transfer. \textbf{a,} The electron spin on the Bloch Sphere. During the state transfer, the $\Omega_{-y}$ spin locking drive (red arrow) defines the high-energy dressed state $\ket*{\tilde{\up}}$ and the low-energy dressed state $\ket*{\tilde{\down}}$ to be along $\ket{-y}$ and $\ket{+y}$, respectively. \textbf{b,} Register manifold reproduced from main text Fig. 1a.   }
    \label{fig:BlochSphere}
\end{figure*}

To achieve accurate spin Rabi frequencies, we include reference Rabi measurements where spin inversion is measured after Rabi drives with $T_\text{Rabi}=0,2,4..80$ ns (Fig. \ref{fig:pulseSequences}c). This allows us to monitor the Rabi frequency on the fly and automatically adjust the Raman power setpoint in case of deviations. 
\\ \\
The ESR probing sequence (Fig. \ref{fig:pulseSequences}d) uses a single electron $\pi$-pulse to estimate the readout counts for a fully inverted electron. Next, it contains two rounds of detuned driving with durations $T_\text{probe}$ and $T_\text{max}-T_\text{probe}$. This ensures a constant duty cycle when $T_\text{probe}$ is scanned from 0 to $T_\text{max}$. Additionally, the two driving steps are repeated with the opposite drive detuning to avoid the build-up of nuclear polarization.
\\ \\
For the NOVEL probe (Fig. \ref{fig:pulseSequences}e), we first include the lock and polarize steps. The probe step consists of a NOVEL drive with spin locking Rabi frequency $\Omega_\text{probe}$ for duration $T_\text{probe}$. We repeat the lock-polarize-probe segment for all values of $T_\text{probe}$ such that a single histogram contains measurements of all drive times for a single $\Omega_\text{probe}$ value. The pulse sequence is then updated with a new value of $\Omega_\text{probe}$. The pulse sequence additionally contains a reference Rabi for long-term stabilization of all electron Rabi frequencies and for establishing the readout counts from $\ketdown$. When recording the thermal NOVEL spectrum (Fig. \ref{fig:NovelComparison}), we keep the polarize step but shift its waveform carrier frequency by 500 MHz (yielding a detuning $\delta/2\pi=\SI{1}{GHz}$), thus suppressing coherent dynamics while maintaining the optical power.
\\ \\
For the magnon Ramsey and memory sequences (Fig. \ref{fig:pulseSequences}f), we include all different combinations of preparation $U_p$ and tomography pulses $U_t$ in the same histogram. Note that $T_\text{store}$ is defined as the delay between the nuclear-resonant spin locking pulses and cannot be reduced below 280 ns due to the intermediate pulses and inter-pulse delays. When varying $T_\text{store}$, we repeat the lock and polarize steps 3 times to compensate for the low duty cycle at long storage times. We include a reference Rabi measurement for the reasons discussed above.
\\ \\
We now elaborate on the effect of the electron control pulses in Fig. \ref{fig:pulseSequences}f. Following previous studies of QDs, we assume a negative electron g-factor such that the bare electron spin state $\ketup$ is on the south pole of the Bloch Sphere (Fig. \ref{fig:BlochSphere}a). 
To realize the SWAP gate, we always apply a $(\pi/2)_x$ rotation before driving around the $-y$ axis. The rotation maps $\ketup$ and $\ketdown$ to the dressed states $\ket*{\tilde{\down}}$ and $\ket*{\tilde{\up}}$ as defined by the $(\Omega)_{-y}$ drive. This mapping constricts the electro-nuclear state to the register manifold (Fig. \ref{fig:BlochSphere}b) during the second half of the protocol where the spin is initialized in $\ketup$ and subjected to a second SWAP. 
Strictly speaking, the second SWAP gate (Fig. \ref{fig:pulseSequences}f) is missing a local electron rotation required to map the dressed states back to the z-basis. We instead include this rotation in the $\hat{U}_t$ rotation. Table \ref{tab:rotations} enumerates the rotation pulses used to realize all 6 input and projective measurements in the quantum tomography (main text Fig. 4b). The fact that the optical readout always prepares $\ketup$ but can only read $\ketdown$ is reflected in the choice of tomography rotations.

\begin{table}[h]
\begin{tabular}{|l|l|l|}
\hline
State & Preparation, $\hat{U}_p$ & Tomography, $\hat{U}_t$ \\ \hline
$+x$   &  $(\pi/2)_{-y}$ & $(\pi/2)_{-y}$            \\ \hline
$-x$   &  $(\pi/2)_y$    & $(\pi/2)_y$             \\ \hline
$+y$   &  $(\pi/2)_{x}$ &  $\mathbb{I}$         \\ \hline
$-y$   &  $(\pi/2)_{-x}$ & $(\pi)_x$           \\ \hline
$+z$   &  $\pi_x$       &  $(\pi/2)_{-x}$          \\ \hline
$-z$   &  $\mathbb{I}$  &  $(\pi/2)_{x}$          \\ \hline
\end{tabular}
\caption{Local electron rotations used to prepare the initial state $\ket{\psi_0}$ and to project the output state onto a target state.}
\label{tab:rotations}
\end{table}

\clearpage
\section{Supplementary measurements}
\subsection{Ramsey interferometry with locked polarisation}
\begin{figure*}[h!]%
\centering
\includegraphics[width=0.6\textwidth]{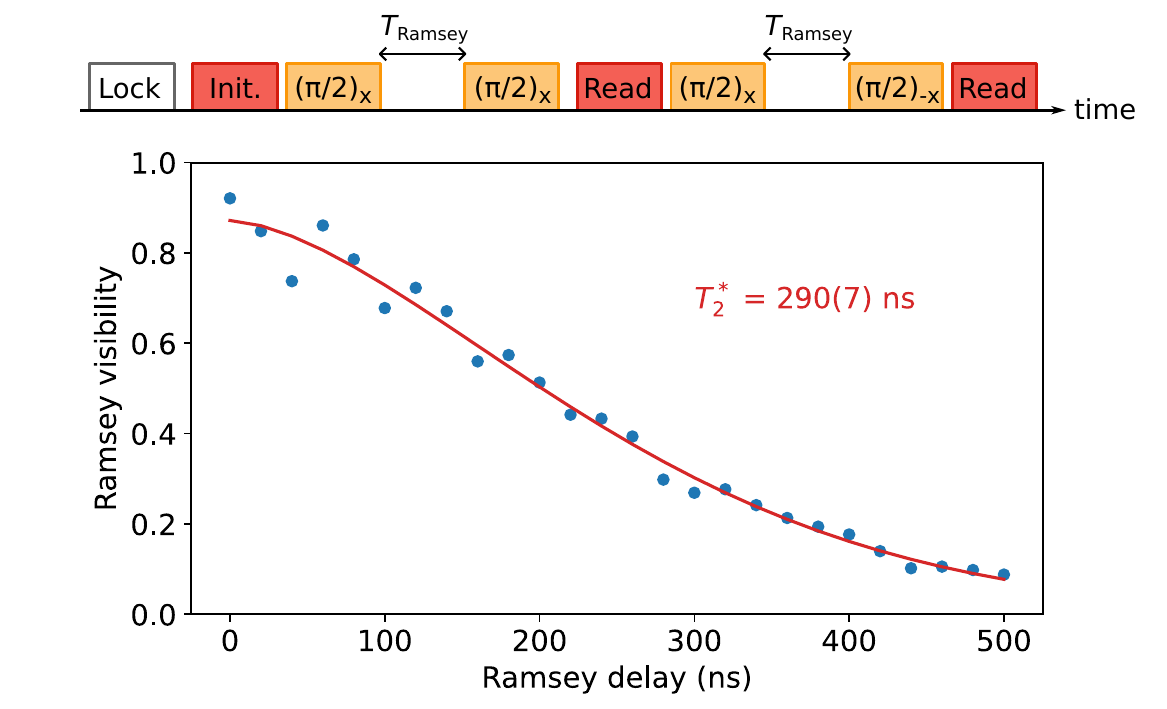}
    \caption{Electron Ramsey measurement preceded by polarisation locking (top inset). The function $v(t)=v_0\times e^{-(t/T_2^*)^\alpha}$ (red curve) is fit to the measured Ramsey visibility (blue points). The extracted fit parameters are $\{v_0, T_2^*, \alpha\}=\{0.87(2), \SI{290(7)}{ns}, 1.62(10)\}$.}
    \label{fig:electronRamsey}
\end{figure*}

Figure \ref{fig:electronRamsey} shows a Ramsey measurement of the electron spin coherence after applying the same nuclear polarisation locking step (Fig. \ref{fig:pulseSequences}a) used in the measurement of the ESR spectrum (Fig. 1d main text). We implement two Ramsey sequences with identical delays but opposite phases for the final $\pi/2$ pulses (Fig. \ref{fig:electronRamsey} inset). The two sequences yield detection counts $n_1$ and $n_2$ from which the visibility $v=(n_1-n_2)/(n_1+n_2)$ is estimated. The estimated dephasing time $T_2^*=\SI{290(7)}{ns}$ corresponds to a Overhauser-induced fluctuation of the qubit detuning $\delta$ with $\sqrt{2}/T_2^*=2\pi\times \SI{0.78(2)}{MHz}$ standard deviation and $2\pi\times\SI{1.83(4)}{MHz}$ FWHM.
\clearpage
\subsection{Interpretation of higher-order magnon sidebands}\label{sec:magnonspectrum}
\begin{figure*}[h!]%
\centering
\includegraphics[width=0.66\textwidth]{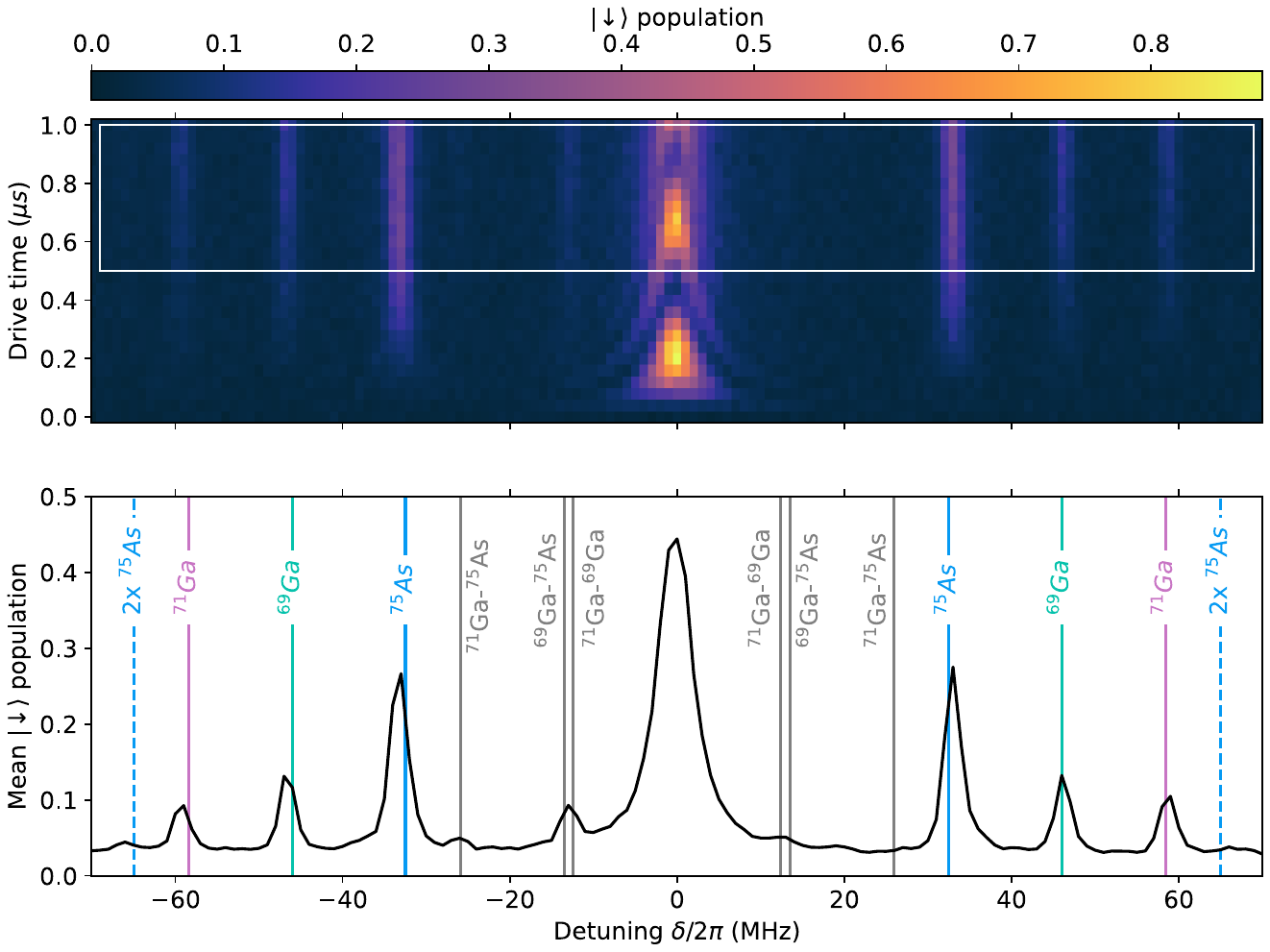}
    \caption{ESR spectrum resulting from an unpolarized nuclear ensemble with locked $I_z\sim 0$ (reproduced main text Fig.1e-f). The additional three-body resonances (gray lines) and the second-order Arsenic transition (blue dashed line) have been marked.}
    \label{fig:magnon_spectrum}
\end{figure*}

We now explain the additional features of the measured ESR spectrum (main text Fig.~1d) which is replotted in Fig. \ref{fig:magnon_spectrum} with additional lines indicating features of interest. 
Firstly, the three-body resonances (between the electron spin and two nuclei) predicted by Eq. (\ref{eq:3bodyv2}) result in ESR peaks at the differences of nuclear Larmor frequencies. For negative drive detunings, we see clear peaks at the expected resonances (solid gray lines). Surprisingly, for positive detunings, the peaks are far less pronounced. This asymmetry may be an artifact of the specific pulse sequence as the three-body sidebands appear symmetric under NOVEL probing (Fig.~\ref{fig:NovelComparison}). Due to the regular spacing of the nuclear Larmor frequencies, we cannot resolve the $^{69}$Ga$-^{75}$As transitions from the $^{71}$Ga$-^{69}$As transitions. 

Secondly, we observe a signature of the second-order arsenic transition (blue dashed line in Fig. \ref{fig:magnon_spectrum}) whereby two excitations are injected into the arsenic ensemble. The large ratio between the first and second-order Arsenic transitions is compatible with a non-collinear interaction dominated by electron g-factor anisotropy, as strain would result in a ratio close to unity (cf. ESR spectrum in Ref. \cite{gangloff_quantum_2019}).

\subsection{Quantum register tomography}
In the quantum tomography in Fig. 4b of the main text, we convert the measured counts to probabilities by normalizing to pairs of orthogonal readouts: The reported probability $p_{a,b}$ of detecting output $\ket{b}$ given input $\ket{a}$ is estimated from $p_{a,b}=n_{a,b}/(n_{a,b}+n_{a,b^\prime})$ where $n$ is the number of detected counts and 
$\braket{b^\prime}{b}=0$. To estimate the register storage fidelity, we first define the contrasts 
\begin{align}
C_\alpha=\frac{1}{2}\left(\frac{n_{+\alpha,+\alpha}-n_{+\alpha,-\alpha}}{n_{+\alpha,+\alpha}+n_{+\alpha,-\alpha}} + \frac{n_{-\alpha,-\alpha}-n_{-\alpha,+\alpha}}{n_{-\alpha,-\alpha}+n_{-\alpha,+\alpha}}\right),\label{eq:contrast}
\end{align}
for $\alpha\in\{x,y,z\}$. Following Ref.\cite{dutt_quantum_2007}, we calculate the storage fidelity $F=(1+\frac{1}{3}(C_x+C_y+C_z))/2$ and propagate the shot noise errors from $n_{a,b}$ onto $F$.

\subsection{Magnon Ramsey}
We perform the magnon Ramsey measurement from main text Fig. 3 using the pulse sequence in Fig. \ref{fig:pulseSequences}f. We utilize four combinations of $U_p$ and $U_t$ to realize the initial states $\ket{\pm x}$ and projective measurements of $\ket{\pm x}$. For each storage time $T_\text{store}$, we extract a $C_x$ contrast using Eq. (\ref{eq:contrast}). 

\begin{figure*}[h!]%
\centering
\includegraphics[width=1\textwidth]{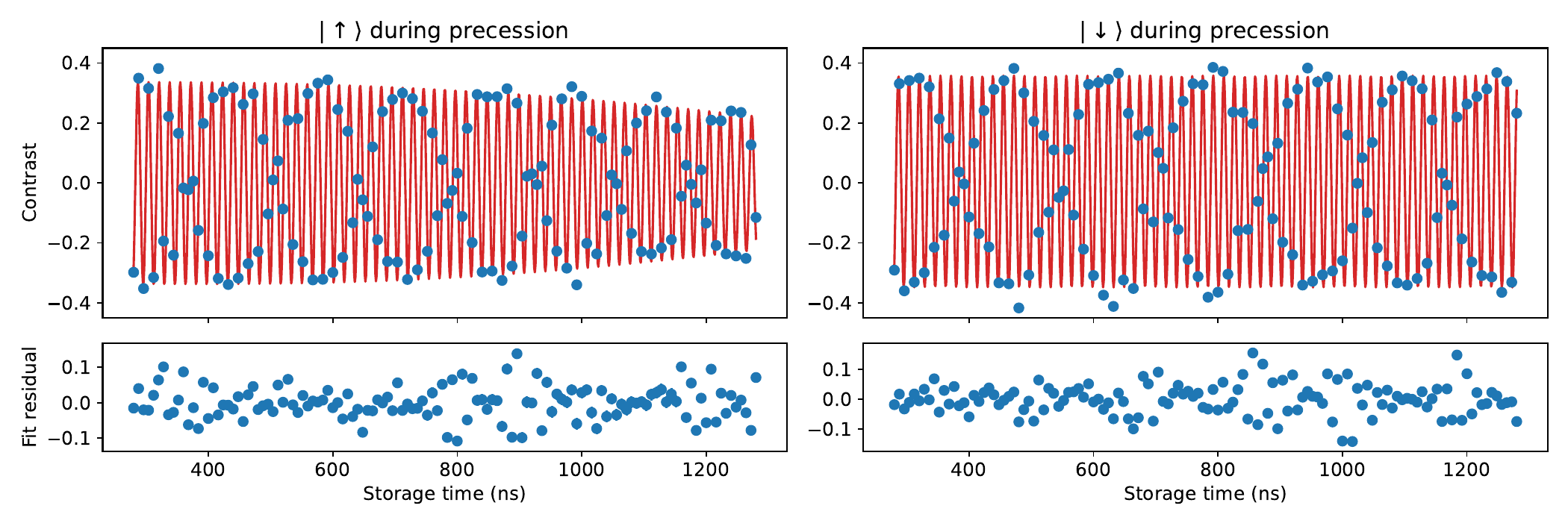}
    \caption{Supplementary measurement of magnon Ramsey. The left and right panels show measurements with the electron in $\ketup$ and $\ketdown$ during precession, respectively. The Ramsey contrast (blue points) is fit with $C_x(t)=C_0\sin(2\pi\nu t+\varphi)e^{-(t/T_{2,\text{mag}}^*)^\alpha}+B$ (red lines). The bottom panels show the fit residuals. 
    The left panel yields the fit parameters $\{C_0, \nu, \varphi, T_{2,\text{mag}}^*, \alpha, B\}=\{0.339(10), \SI{58.810(11)}{MHz}, -0.98(3), \SI{1.4(2)}{\micro s}, 2.6(9), -2(4)\cdot 10^{-3}\}.$
    The right panel yields the fit parameters $\{C_0, \nu, \varphi, T_{2,\text{mag}}^*, \alpha, B\}=\{0.353(8), \SI{59.307(11)}{MHz}, -0.90(3), \SI{1.9(1)}{\micro s}, 3(6), -5(4)\cdot 10^{-3}\}.$
    }
    \label{fig:magnonRamsey}
\end{figure*}

Fig. \ref{fig:magnonRamsey} shows a supplementary magnon Ramsey measurement where $T_\text{store}$ is scanned from 280 ns to 1280 ns in steps of 8 ns. The Nyquist frequency $0.5/(\SI{8}{ns})=\SI{62.5}{MHz}$ is sufficient to resolve the $^{71}$Ga Larmor frequency but leads to the visual illusion of a much slower oscillation. By fitting the data, we obtain the frequency estimates $\nu^\up=\SI{58.810(11)}{MHz}$ and $\nu^\down=\SI{59.307(10)}{MHz}$ for the electron stored in $\ketup$ and $\ketdown$, respectively. The mean frequency $(\nu^\up+\nu^\down)/2=\SI{59.059(8)}{MHz}$ differs from the $\SI{58.810(6)}{MHz}$ estimate reported in the main text. We attribute the difference to a slow discharge of our superconducting magnet coils as the two measurements were taken several days apart. However, the hyperfine-induced frequency difference $\nu^\down-\nu^\up=\SI{0.497(15)}{MHz}$ estimated from Fig. \ref{fig:magnonRamsey} is in statistical agreement with the $\SI{0.500(12)}{MHz}$ value reported in the main text. The longer dephasing time observed in Fig. \ref{fig:magnonRamsey} for $\ketdown$ can be attributed to a partial Knight shift rephasing owing to the included electron inversion.

\subsection{Quantum register storage time estimates}
\begin{figure*}[h!]%
\centering
\includegraphics[width=1\textwidth]{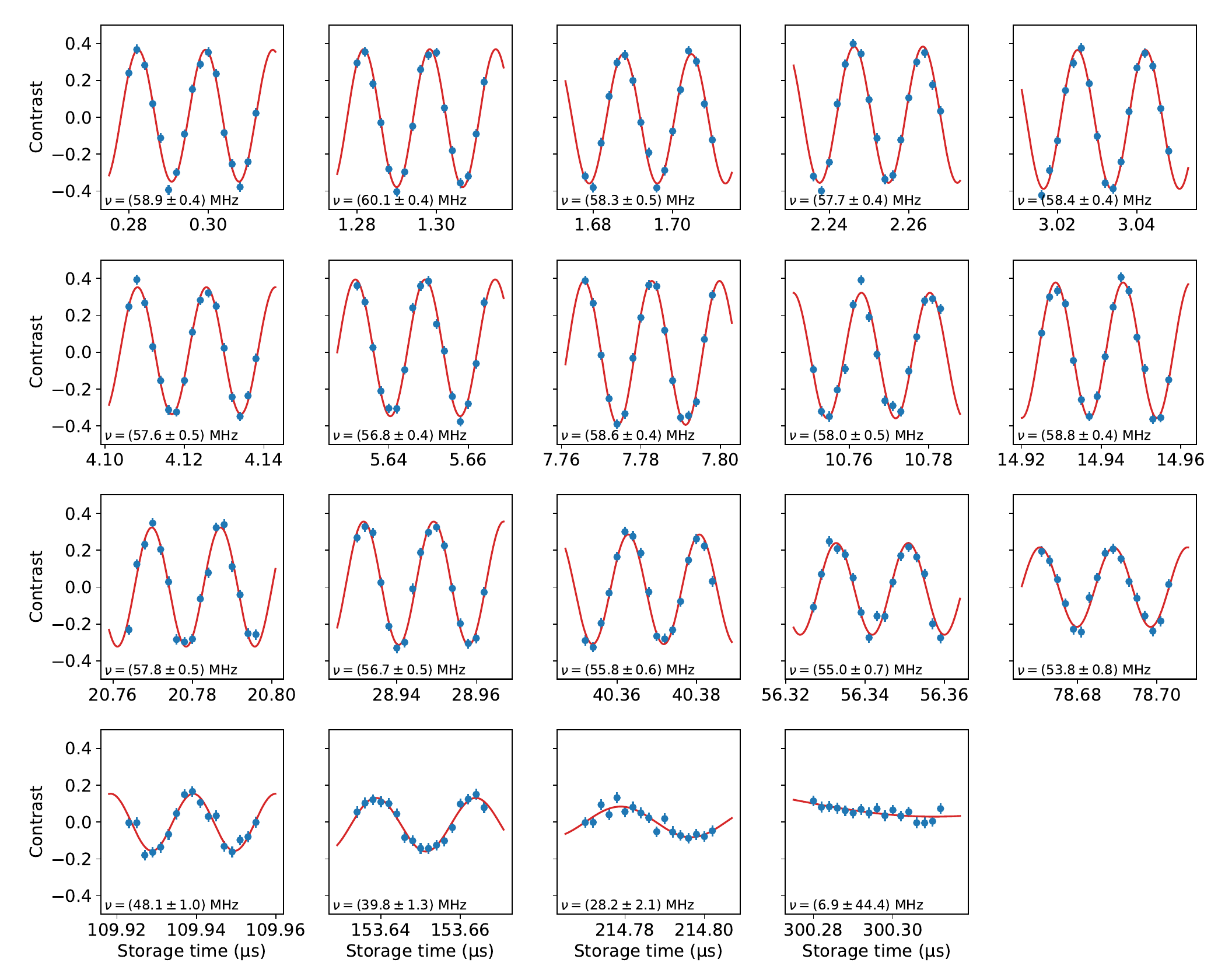}
    \caption{Measured segments (blue dots) of magnon Ramsey used to estimate the storage time of $\ket{x}$. Red curves indicate fits. The fitted magnon precession frequency $\nu$ is given inside each subpanel.  
    }
    \label{fig:memory_subfits}
\end{figure*}

\begin{figure*}[h!]%
\centering
\includegraphics[width=0.5\textwidth]{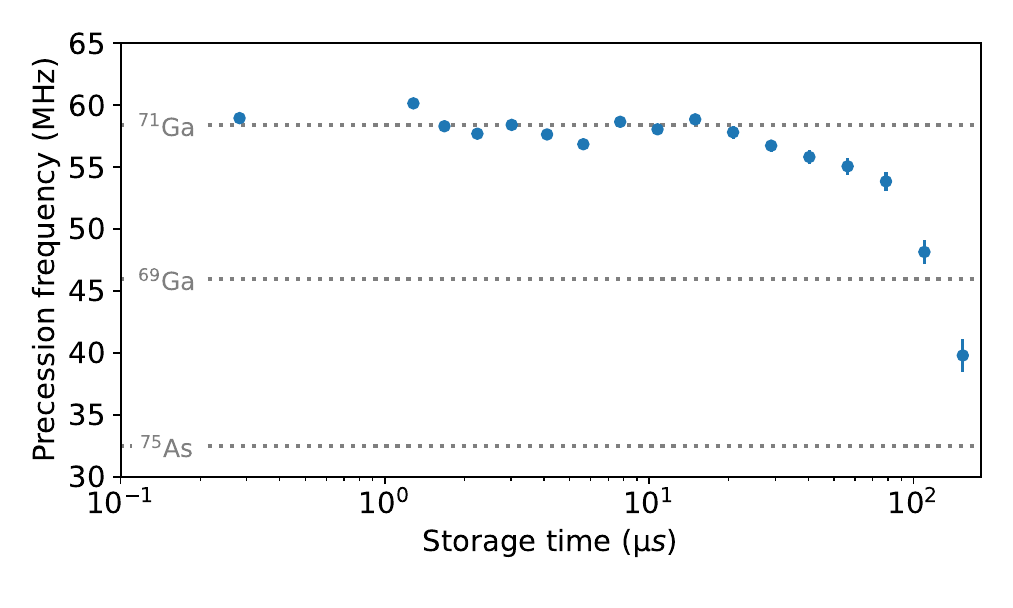}
    \caption{Extracted magnon precession frequencies from Fig. \ref{fig:memory_subfits}. The dashed lines indicate the Larmor frequencies of the three nuclear species.
    }
    \label{fig:memory_freqdrift}
\end{figure*}

\begin{figure*}[h!]%
\centering
\includegraphics[width=0.5\textwidth]{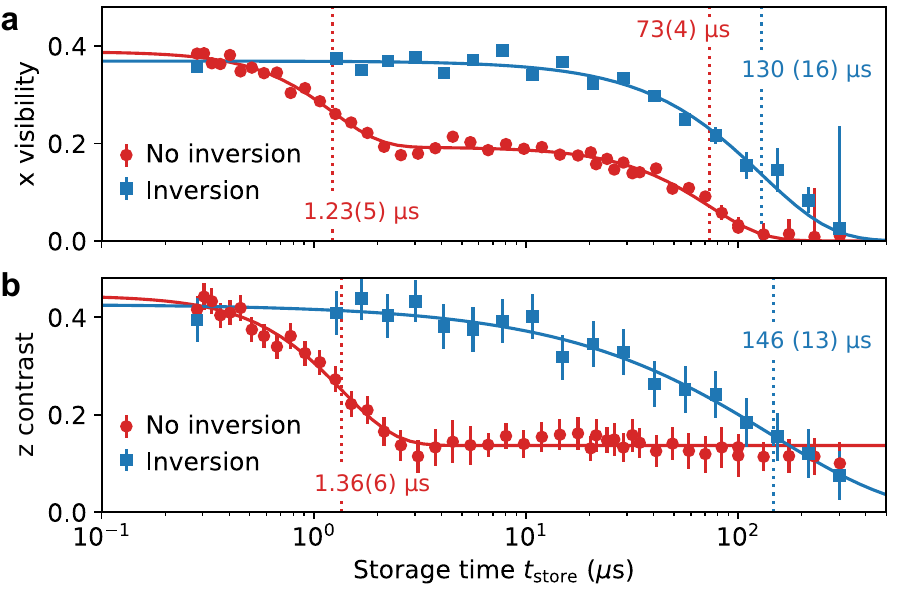}
    \caption{
    \textbf{a,} Ramsey visibility of an $\ket{x}$ input state (reproduced from main text Fig. 4c). 
    \textbf{b,} Storage contrast of the input state $\ket{\psi_0}=\ket{\pm z}$. Red data points are fit with $C(t)=C_0 e^{-(t/T_a)^\alpha}+B$ yielding $\{C_0,T_a,\alpha,B\}=\{0.308(15),\SI{1.36(6)}{\micro s},1.7(2),0.137(3)\}$ where $B$ is an empirical parameter describing the residual visibility.  Including the electron inversion yields the blue points which are fit with $C(t)=C_0 e^{-(t/T_a)^\alpha}$ yielding $\{C_0,T_a,\alpha\}=\{0.427(13),\SI{146(13)}{\micro s},0.74(9)\}$. The dashed lines indicate $T_a$.
    }
    \label{fig:xz_storage}
\end{figure*}

Fig. \ref{fig:memory_subfits} shows the short segments of magnon Ramsey from which we extract the storage time-dependent Ramsey visibility in the presence of an electron spin inversion pulse. 
At each storage time, we store and retrieve the qubit state $\ket{\pm x}$, perform projective readout of $\ket{\pm x}$ and estimate a $C_x$ contrast following Eq. (\ref{eq:contrast}).
The contrast is fit with the model $C_x(t)=C_0\sin(2\pi\nu t+\varphi)+B$ where $C_0$ is the Ramsey visibility, $\nu$ is the precession frequency, $\varphi$ is the fringe phase and $B$ is an empirical background. All four parameters are kept free across all datasets.
As the storage time is increased, in addition to a decay $C_0$, we observe a decay in the precession frequency $\nu$ as evident from the final fits in Fig. \ref{fig:memory_subfits}.
This frequency shift is plotted in Fig. \ref{fig:memory_freqdrift}. We do not currently understand the cause of this drift and have not identified any systematic errors in the synthesis of the control pulses determining the storage time.
We note the stored qubit state persists in the nuclear ensemble despite the precession slowdown.
\clearpage
Fig. \ref{fig:xz_storage}a shows the register storage time for a $\ket{\pm x}$ input state (reproduced from main text Fig. 4c) while Fig. \ref{fig:xz_storage}b shows the storage time of a $\ket{\pm z}$ input state.
For $\ket{\pm z}$, each data point corresponds to a single storage time where we estimate the $C_z$ contrast using Eq. (\ref{eq:contrast}). Fig. \ref{fig:xz_storage}b shows a fast decay of contrast in the absence of an electron inversion (red dots) similar to the case of $\ket{\pm x}$ storage. We again attribute this decay to the Knight field inhomogeneity which breaks the conservation of $j$~\cite{zaporski_many-body_2023} and inhibits a perfect magnon retrieval. In contrast to $\ket{\pm x}$ storage, we observe a residual z-visibility, which we attribute to a relaxation of nuclear polarization ($T_1$) slower than the nuclear dephasing time $T_2^*$. Adding an electron inversion (blue squares) similarly prolongs the storage time.

\clearpage
\section{Estimation of the number of nuclei}\label{sec:numberOfNuclei}
The spatially varying wavefunction of the confined electron results in a gradient in the collinear and non-collinear hyperfine couplings. However, to simulate the system dynamics, we assume $N$ nuclei with a uniform hyperfine coupling to the electron. We estimate this effective $N$ from the differential Knight shift measured in the main text. This estimator is however biased by the fact that nuclei with strong hyperfine couplings contribute a bigger amplitude to the collective excitation while simultaneously experiencing a larger Knight shift. Assuming that the nuclear dark state represents a fully polarized nuclear ensemble, the energy difference between zero nuclear excitations $\ket{0}$ and a single excitation $\ket{1}$ is given by:
\begin{align}
    \Delta E /\hbar&= \bra{1}\hat{H}\ket{1} - \bra{0}\hat{H}\ket{0} = \bra{0} \Phi^- \hat{H} \Phi^+ \ket{0} - 0 \\
    &= \frac{1}{\sum_i a_i^2} \bra{0} \sum_i a_i I_-^{(i)} \sum_j (\omega_n + S_z a_j)I_z^{(j)} \sum_k a_k I_+^{(k)} \ket{0}\\
    &=\omega_n+S_z\frac{\sum_ia_i^3}{\sum_i a_i^2},\label{eq:knightshift}
\end{align}
where the sums run over individual nuclei and $\Phi^+ = \sum_ia_iI_+^{(i)}/\sqrt{\sum_i a_i^2}$ is the normalized collective raising operator \cite{Hu_entangled_2015}. 
The observed Knight shift thus depends on the second and third moments of the distribution of hyperfine couplings. The hyperfine coupling of a nucleus located at position $\mathbf{r}_i$ is given by $a_i=A|\psi(\mathbf{r}_i)|^2$ where $A$ is the hyperfine material constant and $|\psi(\mathbf{r})|^2$ is the normalized electron envelope wavefunction \cite{urbaszek_nuclear_2013}. We now assume a Gaussian envelope
\begin{align}
    |\psi(\mathbf{r})|^2=\frac{e^{-r^2/\sigma^2}}{\sigma^3\pi^{3/2}},
\end{align}
where $r=|\mathbf{r}|$ and $\sigma$ is the characteristic width. This satisfies the normalisation $\int_V dV|\psi(\mathbf{r})|^2 = \int_0^\infty dr |\psi(\mathbf{r})|^2r^24\pi=1$. 
Converting the sums over nuclei to integrals, we obtain the moments
\begin{align}
    \sum_i a_i &\rightarrow \int_0^\infty dr A|\psi(\mathbf{r})|^2 r^2 4\pi = A,\\
    \sum_i a_i^2 &\rightarrow  \int_0^\infty dr (A|\psi(\mathbf{r})|^2)^2 r^2 4\pi = A^2/(\sigma^3\sqrt{8}\pi^{3/2}),\label{eq:asquaredinhom}\\
    \sum_i a_i^3 &\rightarrow  \int_0^\infty dr (A|\psi(\mathbf{r})|^2)^3 r^2 4\pi = A^3/(\sigma^6\sqrt{27}\pi^3).\label{eq:cubedinhom}
\end{align}
Note that a Gaussian distribution with different widths along the $x,y$ and $z$ axes will result in the same moments under the replacement $\sigma^3\rightarrow\sigma_x\sigma_y\sigma_z$, where $\sigma_\alpha$ is the width along axis $\alpha$. The second moment is the first non-trivial moment and influences the thermal electron $T_2^*$ which is often used to estimate the number of nuclei \cite{zaporski_ideal_2023,gangloff_quantum_2019}. We therefore use this moment to define an effective number of nuclei. A uniform collection of $N$ nuclei with $a_i=A/N$ results in 
\begin{align}
    \sum_i^N a_i^2 = \sum_i^N\left(\frac{A}{N}\right)^2 = \frac{A^2}{N}.\label{eq:asquaredhom}
\end{align}
Equating Eq. (\ref{eq:asquaredhom}) and Eq. (\ref{eq:asquaredinhom}) then yields
\begin{align}
    N = \sigma^3\sqrt{8}\pi^{3/2}.\label{eq:Neff}
\end{align}
Substituting this result into Eqs. (\ref{eq:knightshift},\ref{eq:asquaredinhom},\ref{eq:cubedinhom}) leads to 
\begin{align}
\Delta E/\hbar -\omega_n= \frac{\sum_ia_i^3}{\sum_i a_i^2} = \frac{A}{\sigma^3}\sqrt{\frac{8}{27\pi^3}} = \frac{8}{\sqrt{27}}\frac{A}{N}\approx 1.54\frac{A}{N}.
\end{align}
As expected, the measurable Knight shift is greater than the Knight shift from a uniformly coupled ensemble. The measured $^{71}$Ga Knight shift is given by
\begin{align}
    \delta\nu = \frac{1.54\times c_{71}(A_{71}/2\pi)}{N_{71}},
\end{align}
where $c_{71},A_{71}$ and $N_{71}$ are the abundance, the hyperfine constant and the number of nuclei of $^{71}$Ga, respectively.
Using the measured $\delta\nu=\SI{0.500(12)}{MHz}$ and the material constants in table \ref{tab:nuclearconstants}, we estimate $N_{71}=1.35(3)\cdot 10^4$. The number of effective nuclei across all species is then $N_\text{tot}=2\times N_{71}/c_{71}=6.84(16)\cdot 10^4$, where the factor 2 accounts for the presence of arsenic and gallium in the unit cell. Note that the quoted errors on $N_\text{tot}$ only reflect the statistical fitting error on the $\delta\nu$ estimate.

\section{System hamiltonian}
We consider the case discussed in Ref. \cite{botzem_quadrupolar_2016} where an electron g-factor anisotropy tilts the electron quantization axis by angle $\phi$ away from the external field $\mathbf{B}$, giving rise to the Hamiltonian 
\begin{align}
    \hat{H} = \omega_e \cos(\phi) \hat{S}_z + \omega_e \sin(\phi)\hat{S}_x + \sum_i\omega_i\hat{I}_z^{(i)} + \sum_ia_i\mathbf{S}\cdot\mathbf{I}^{(i)},
\end{align}
where $\omega_e$ is the electron Zeeman splitting, $\hat{\mathbf{S}}$ is the electron spin operator, and $\hat{\mathbf{I}}^{(i)}$, $\omega_i$, $a_i$ denote the spin operator, Larmor frequency and hyperfine coupling of the i'th nucleus, respectively.
In this coordinate system, the $y$-axis is the QD growth direction. 
For a B-field 45$\degree$ in-between the $[110]$ and $[\Bar{1}10]$ crystallographic axes, the electron quantisation axis tilt $\phi$ is given by 
\begin{align}
    \tan(\phi)=\frac{g_{110}-g_{\bar{1}10}}{g_{110}+g_{\bar{1}10}},
\end{align}
where the above g-factors are along the crystallographic axes in which the g-tensor is approximately diagonal \cite{botzem_quadrupolar_2016}.
We first diagonalize the electron Zeeman interaction with the unitary transform $\hat{U}=e^{i\phi S_y}$ which leads to 
\begin{align}
    \hat{H}^\prime = \hat{U}\hat{H}\hat{U}^\dag =\omega_e \hat{S}_{z^\prime} + \sum_i\omega_i \nuc{z} + \sum_i a_i\left[\hat{S}_{x^\prime}(\cos(\phi)\nuc{x}-\sin(\phi)\nuc{z})+\hat{S}_{y^\prime}\nuc{y}+\hat{S}_{z^\prime}(\cos(\phi)\nuc{z}+\sin(\phi)\nuc{x})\right]\label{eq:H2}
\end{align}
where the primed coordinates indicate a coordinate system co-aligned with the electron quantization axis. The non-secular terms containing $I_xS_{x^\prime}$ and $I_yS_{y^\prime}$ are suppressed to first order by the qubit splitting $\omega_e$. By performing a Schrieffer-Wolff transformation \cite{bravyi_schriefferwolff_2011} with small expansion parameter $a^2/\omega_e$, we thus obtain
\begin{align}
    \hat{H}^\prime = \omega_e \hat{S}_{z^\prime}  + \sum_i\omega_i \nuc{z} +\hat{S}_{z^\prime}\sum_i\left[ a_\parallel^{(i)} \nuc{z}+a_\perp^{(i)}(\nuc{+}+\nuc{-})/2\right] + \hat{H}_\text{3b},\label{eq:HMain}\\
    \hat{H}_\text{3b} =\hat{S}_{z^\prime}\sum_{i,k}\frac{a_ia_k}{4\omega_e}\left[(1+\cos(2\phi))\hat{I}_x^{(i)}\hat{I}_x^{(k)} + 2\hat{I}_y^{(i)}\hat{I}_y^{(k)} + (1-\cos(2\phi))\hat{I}_z^{(i)}\hat{I}_z^{(k)}\right],\label{eq:3body}
\end{align}
where we defined the collinear and non-collinear couplings $a_\parallel^{(i)} = a_i\cos(\phi)$ and $a_\perp^{(i)} = a_i\sin(\phi)$, respectively, and we additionally used $\hat{I}_x=(\hat{I}_+ + \hat{I}_-)/2$. 
The first three terms in Eq. (\ref{eq:HMain}) lead to Eq. (1) from the main text. $\hat{H}_\text{3b}$ represents weaker three-body interactions where the electron spin couples to pairs of nuclei. In terms of nuclear raising and lowering operators, this term becomes
\begin{align}
     \hat{H}_\text{3b}=\hat{S}_{z^\prime}\sum_{i,k}\frac{a_ia_k}{4\omega_e}\left[
     \frac{3+\cos(2\phi)}{4}(\hat{I}_+^{(i)}\hat{I}_-^{(k)} + \hat{I}_-^{(i)}\hat{I}_+^{(k)})
     +\frac{\cos(2\phi)-1}{4}(\hat{I}_+^{(i)}\hat{I}_+^{(k)} + \hat{I}_-^{(i)}\hat{I}_-^{(k)}) 
     + (1-\cos(2\phi))\hat{I}_z^{(i)}\hat{I}_z^{(k)}
     \right].\label{eq:3bodyv2}
\end{align}
The first term in Eq. (\ref{eq:3bodyv2}) results in electron-mediated flip-flops between different nuclei and manifests in peaks in the ESR spectrum at the differences of nuclear Larmor frequencies (section \ref{sec:magnonspectrum}). The second term represents a second-order process where two nuclear excitations are created at once. It is however very weak due to the quadratic $\phi$-dependence and is not considered further. The final term represents a similarly weak renormalization of the electron hyperfine shift.

\subsection{Dressed state picture}
We now consider the rate of magnon activation in the presence of the first-order hyperfine coupling and electron drive as described by
\begin{align}
    \hat{H} = \delta \hat{S}_z + \Omega \hat{S}_x + \omega_n \hat{I}_z + \hat{S}_z(a_\parallel \hat{I}_z+a_\perp \hat{I}_x),
\end{align}
where $\delta$ is the drive detuning, $\Omega$ is the spin Rabi frequency and we have dropped the primed electron coordinates and additionally utilize the collective nuclear operators.  As we can absorb the hyperfine shift $a_\parallel I_z$ of the initial state into the drive detuning $\delta$, this term only contributes an electron-dependent Knight shift to the transition energy between the nuclear states $\ket{I_z}$ and $\ket{I_z\pm 1}$.
We transform to the dressed electron states using the transformation 
\begin{align}
\hat{U}_d = \begin{bmatrix}
\sin(\theta) & \cos(\theta)\\
\cos(\theta) & -\sin(\theta)
\end{bmatrix}_\text{electron}\otimes \mathbb{I}_\text{nuc}\;, \; \tan(2\theta)=-\frac{\Omega}{\delta},
\end{align}
giving rise to the dressed state Hamiltonian keeping $z$ as the quantization axis:
\begin{align}
\tilde{H}=\chi\tilde{S}_z + \omega_n \hat{I}_z - \frac{\delta}{\chi}\left(\tilde{S}_z+\frac{\Omega}{\delta}\tilde{S}_x\right)\left(a_\parallel \hat{I}_z + \frac{a_\perp}{2}(\hat{I}_+ + \hat{I}_-)\right), \label{eq:dressedH1}
\end{align}
where $\chi=\sqrt{\Omega^2+\delta^2}$ is the generalised Rabi frequency. 
The matrix element signifying an electro-nuclear swap is 
\begin{align}
\frac{\Omega_{mag}^+}{2} &= \left|\bra{j,m+1,\tilde{\down}} \tilde{H} \ket{j,m,\tilde{\up}}\right| = \frac{a_\perp\Omega}{4\chi}\bra{j,m+1}\hat{I}_+\ket{j,m},\label{eq:MagnonRate}
\end{align}
with a collective enhancement given by
\begin{align}
\bra{j,m+1}\hat{I}_+\ket{j,m} = \sqrt{j(j+1)-m(m+1)}.\label{eq:LadderMatrixElement}
\end{align}
In the detuned driving limit, $\delta\gg \Omega$, the rate reduces to $\Omega^+_\text{mag}= \frac{\Omega a_\perp}{2\delta}|\hat{I}_+|$ which contains a $1/\delta$ drop-off. Inspecting the diagonal elements of Eq. (\ref{eq:dressedH1}), the resonance condition is $\chi=\omega_n+ a_\parallel S_0$ where $a_\parallel S_0$ is the differential Knight shift between the final and initial state given an initial electron spin $z$-projection $S_0$. As $\chi\approx\delta$, the magnon transition inherits the inhomogeneous electron linewidth set by $T_2^*$.

In the NOVEL scheme used for polarization and state transfer, $\delta=0$ and $\chi=\Omega$. This has the benefit that $\chi$ is first-order insensitive to $\delta$ fluctuations related to the finite $T_2^*$. In this case, the dressed electron states are equal superpositions of the bare states 
and the Hamiltonian reduces to
\begin{align}
    \tilde{H}=\Omega\tilde{S}_z + \omega_n \hat{I}_z - a_\parallel\tilde{S}_x  \hat{I}_z + \frac{a_\perp}{2}\tilde{S}_x(\hat{I}_+ + \hat{I}_-). \label{eq:dressedH3}
\end{align}
The first two terms dictate the resonance condition $\Omega=\omega_n$. The third term in Eq. (\ref{eq:dressedH3}) represents a weak coupling through the collinear term. As we can absorb $\hat{I}_z$ of the initial state into $\delta$, this term of magnitude $a_\parallel$ becomes insignificant. Instead, the dynamics are dominated by the last term in Eq. (\ref{eq:dressedH3}) resulting in the magnon Rabi frequency
\begin{align}
\frac{\Omega_\text{mag}^+}{2} = \left|\bra{j,m+1,\tilde{\down}} \tilde{H} \ket{j,m,\tilde{\up}}\right| = \frac{a_\perp}{4}\bra{j,m+1}\hat{I}_+\ket{j,m}.
\end{align}
For a dark state $m=-j$, this results in the rate $\Omega^+_\text{mag}=\frac{a_\perp}{2}\bra{j,-j+1}\hat{I}_+\ket{j,-j}=a_\perp\sqrt{j/2}$ following Eq. (\ref{eq:LadderMatrixElement}).
\clearpage

\section{Monte Carlo Simulation}
We now describe the Monte Carlo simulation used to simulate the NOVEL spectra and the quantum state transfer. Simulating the large nuclear ensembles is enabled by working in truncated subspaces of the collective nuclear basis with initial nuclear spin states sampled at random. 

\subsection{Coherent evolution}
The simulation Hilbert space $\mathds{C}^2\otimes\mathds{C}^3\otimes\mathds{C}^3\otimes\mathds{C}^5$ consists of the electron spin, the two Gallium ensembles, and the Arsenic ensemble. 
For the gallium ensembles, we include the collective states $\{\ket{j,m+1},\ket{j,m},\ket{j,m-1}\}$, where $j$ is the spin length and $m$ is the spin z-projection. For arsenic, we include 
$\{\ket{j,m+2}$, $\ket{j,m+1}$, $
\ket{j,m}$,  $\ket{j,m-1}$, $\ket{j,m-2}\}$ in order to reproduce the weak second-order arsenic transition. We apply the Hamiltonian in Eq. (\ref{eq:HMain}) but only include the first three-body term in Eq. (\ref{eq:3bodyv2}) to arrive at 
\begin{align}
\hat{H}_\text{mc} = 
\sum_i^N \omega_i \hat{I}_i^z 
+ \hat{S}_z\sum_i^N \frac{a_\perp^{(i)}}{2}(\nuc{+}+\nuc{-}) 
+ \hat{S}_z\sum_{i,j}^N \frac{a_\parallel^{(i)}a_\parallel^{(j)}}{4\omega_e}(\hat{I}_+^{(i)}\hat{I}_-^{(j)} 
+ \hat{I}_-^{(i)}\hat{I}_+^{(j)})
+\hat{H}_{2\times\text{As}} + \hat{H}_\text{drive}.\label{eq:H_mc}
\end{align}
In the truncated Hilbert space, we implement the gallium raising and lowering operators in the matrix form 
\begin{align}
    \nuc{+} = \left(
    \begin{array}{ccc}
0 & a & 0 \\ 
0 & 0 & b \\ 
0 & 0 & 0
\end{array} 
\right),\\
a = \sqrt{j_i(j_i+1)-m_i(m_i+1)},\\
b = \sqrt{j_i(j_i+1)-(m_i-1)(m_i)},\\
\nuc{-}=(\nuc{+})^\dag,
\end{align}
where $j_i$ and $m_i$ relate to the initial state $\ket{j_i,m_i}$ for species $i$. For arsenic, $\nuc{\pm}$ is similarly implemented with a 5x5 matrix. 

We additionally wish to reproduce the nuclear resonance at twice the arsenic Larmor frequency observed in the ESR spectrum (Fig. \ref{fig:magnon_spectrum}) and in NOVEL driving (main text Fig. 2b). The double-magnon transition rates expected from our device strain (section \ref{sec:strainEstimation}) and the three-body term (Eq. \ref{eq:3bodyv2}) are however too small to reproduce the observed rate. As this simulation is intended to reproduce the errors on state transfer from overlapping nuclear resonances, it is sufficient to incorporate an empirical Hamiltonian term
\begin{align}
\hat{H}_{2\times\text{As}} = \eta_{2\times \text{As}}\hat{S}_z [(I_+^\text{As})^2+(I_-^\text{As})^2], 
\end{align}
where $\eta_{2\times\text{As}}$ is estimated to be approximately $2\pi\times11$ Hz from polarized NOVEL spectra (Fig. \ref{fig:NovelComparison}).

Finally, the time-dependent electron drive in the rotating frame is given by 
\begin{align}
\hat{H}_{drive}(t) = \delta \hat{S}_z + \Omega_x(t)\hat{S}_x+\Omega_y(t)\hat{S}_y,
\end{align}
where the timescale represented by t is much longer than electron precession time. Note that Eq. (\ref{eq:H_mc}) does not include the collinear hyperfine interaction $a_\parallel \hat{S}_z\hat{I}_z$ which limits the electron $T_2^*$. We instead implement this effect by randomly sampling $\delta$ from a Gaussian distribution with $\sigma=\sqrt{2}/T_2^*$ standard deviation.
\subsection{Nuclear state sampling}\label{sec:thermalSampling}
We now describe how to sample $\ket{j,m}$ for a thermal state. 
We note that an equal mixture of $N$ spin-1/2s is completely diagonal in the $\ket{j,m}$ basis with a joint probability mass function given by \cite{wesenberg_mixed_2002}
\begin{align}
    p_{j,m} = \frac{2j+1}{j_0+j+1}\binom{2j_0}{j_0+j}p^{j_0+m}(1-p)^{j_0-m},\label{eq:spinhalfpdf}
\end{align}
where $j_0=N/2$ is the maximal spin length and $p$ is the probability of a single spin being excited. The temperatures and magnetic fields considered in this work correspond to the infinite temperature limit with no thermal spin inversion and $p=1/2$. This simplifies Eq. (\ref{eq:spinhalfpdf}) which no longer depends on $m$. The marginal distribution $p_j$ is then given by
\begin{align}
    p_j = p_{j,m} \times (2j+1) = \frac{(2j+1)^2}{j_0+j+1}\binom{2j_0}{j_0+j}\label{eq:jmarginal},
\end{align}
as the angular momentum $j$ accommodates $2j+1$ equally likely polarisations. The strategy for sampling $\ket{j,m}$ is now clear: Sample $j$ from Eq. (\ref{eq:jmarginal}) and sample $m$ from the uniform $-j..j$ distribution.

For the spin-3/2 nuclei considered in this work, explicitly calculating $p_{j,m}$ for $N\sim 10^5$ is computationally challenging \cite{wesenberg_mixed_2002}. Instead, 
by working with a fully mixed state of many spins, we note that $I_x$, $I_y$, and $I_z$ are largely uncorrelated such that $I^2= I_x^2+I_y^2+I_z^2\approx 3I_z^2$. For $N$ thermal nuclei of spin $I$, the variance $\expval{m^2}-\expval{m}^2=\frac{1}{3}NI(I+1)$ equals $N/4$ and $5N/4$ for $I=1/2$ and $I=3/2$, respectively. Thus, $5N$ spin-1/2s reproduce the same statistical distribution in $m$ (and by extension $j$) as $N$ spin-3/2s. This allows us to reuse the spin-1/2 sampling strategy by simply scaling the number of nuclei.
\subsection{Non unitary dynamics}
We incorporate incoherent electron spin flips by solving the master equation
\begin{align}
\frac{d}{dt}\hat{\rho}=-i[\hat{H}_\text{mc},\hat{\rho}] + \sum_{j} \left[\hat{C}_j\hat{\rho} \hat{C}_j^\dag - (\hat{C}_j^\dag \hat{C}_j \hat{\rho} + \hat{\rho} \hat{C}_j^\dag \hat{C}_j)/2\right],
\end{align}
where $\hat{\rho}$ is the electro-nuclear density matrix. We include the collapse operators $\hat{C}_1=\sqrt{\kappa}\ketbra{\up}{\down}\otimes\mathds{I}_\text{nuc}$ and $\hat{C}_2=\sqrt{\kappa}\ketbra{\down}{\up}\otimes\mathds{I}_\text{nuc}$, where $\mathds{I}_\text{nuc}$ is the identity operator on all nuclear ensembles. Following the observations in Ref. \cite{bodey_optical_2019}, we take the spin-flip rate $\kappa$ to be proportional to the drive Rabi frequency, $\kappa=|\Omega|/(2Q)$. For an undriven electron, this results in a relaxation time $T_1=Q(2\pi/\Omega)$. We numerically integrate the master equation using the QuTip Python library. We note that spin flips can alternatively be incorporated with Monte Carlo wavefunctions which offer significant computational performance gains at the cost of numerical noise. 

\subsection{Q-factor estimation}
\begin{figure*}[h!]%
\centering
\includegraphics[width=0.8\textwidth]{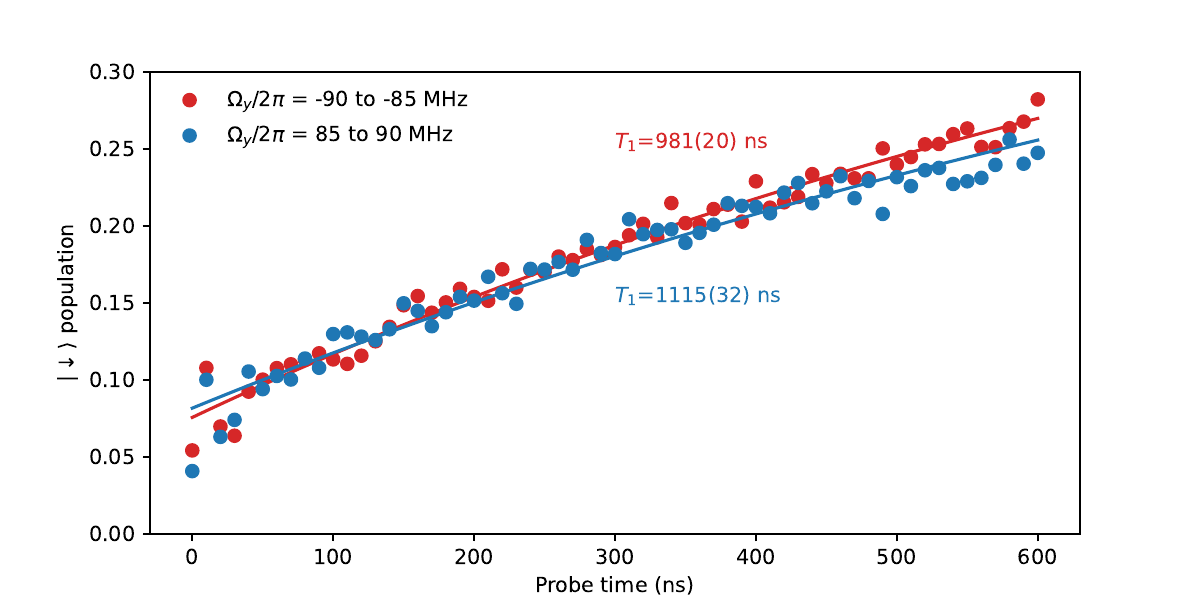}
    \caption{Estimation of electron $T_1$ time. The dots denote the measured NOVEL signals for an unpolarized ensemble (Fig. \ref{fig:NovelComparison}) averaged across drive frequencies of -85 to -90 MHz (red dots) and 85 to 90 Mhz (blue dots). Fits (solid curves) use the model $p_\up(t)=(0.5-p_0)\cdot(1-e^{(-t/T_1)})+p_0$ where $p_\up$ is the electron $\ketup$ population and $p_0$ is the initial electron inversion. This model ensures electron spin depolarization at long drive times, i.e. $p_\up(t\rightarrow\infty)=0.5$.}
    \label{fig:NovelQ}
\end{figure*}

For an electron driven in a spin-locking configuration away from nuclear resonance, the master equation results in a prolonged $T_1=2\times Q(2\pi/\Omega)$ relaxation time. We use this fact to estimate $Q$ from the spin-locking data. Figure \ref{fig:NovelQ} shows $T_1$-fits to the NOVEL signal. By averaging the extracted $T_1$ estimates, we estimate $Q=46$.

\subsection{NOVEL probe simulation and anisotropy estimation}\label{sec:NovelSim}
\begin{figure*}[h!]%
\centering
\includegraphics[width=1\textwidth]{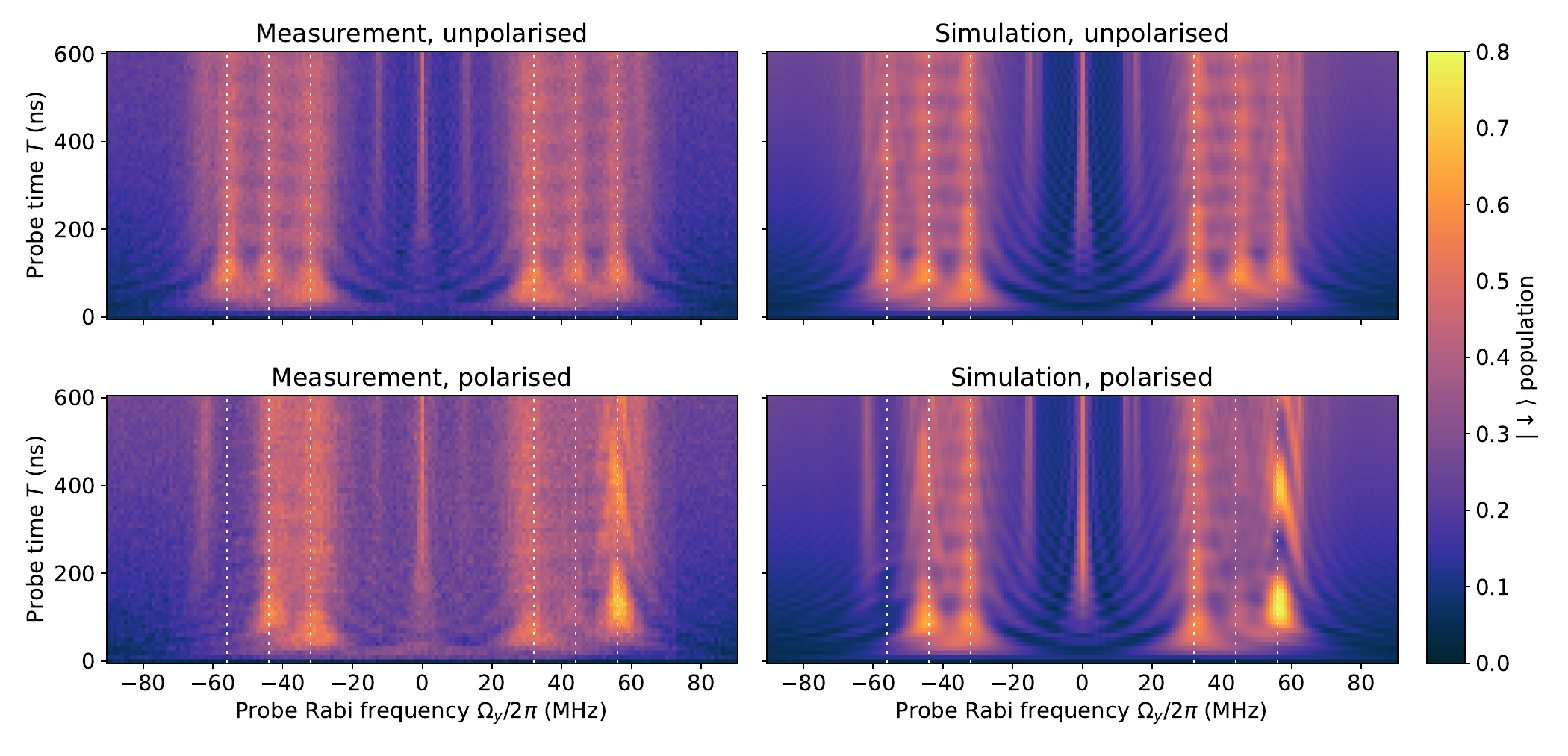}
    \caption{Measurements and simulations of NOVEL spectra for an unpolarized and polarized nuclear ensemble. The bottom row is identical to Fig.~2b and Fig.~2d in the main text. The vertical dashed lines correspond to $\pm 32$, $\pm 44$, and $\pm 56$ MHz. For the simulations, we average over 100 randomly sampled initial nuclear states. The simulations include imperfections from the overlap of nuclear species, electron spin inhomogeneous broadening, incoherent electron spin-flips, and electron spin initialization.   }
    \label{fig:NovelComparison}
\end{figure*}

\begin{figure*}[h!]%
\centering
\includegraphics[width=1\textwidth]{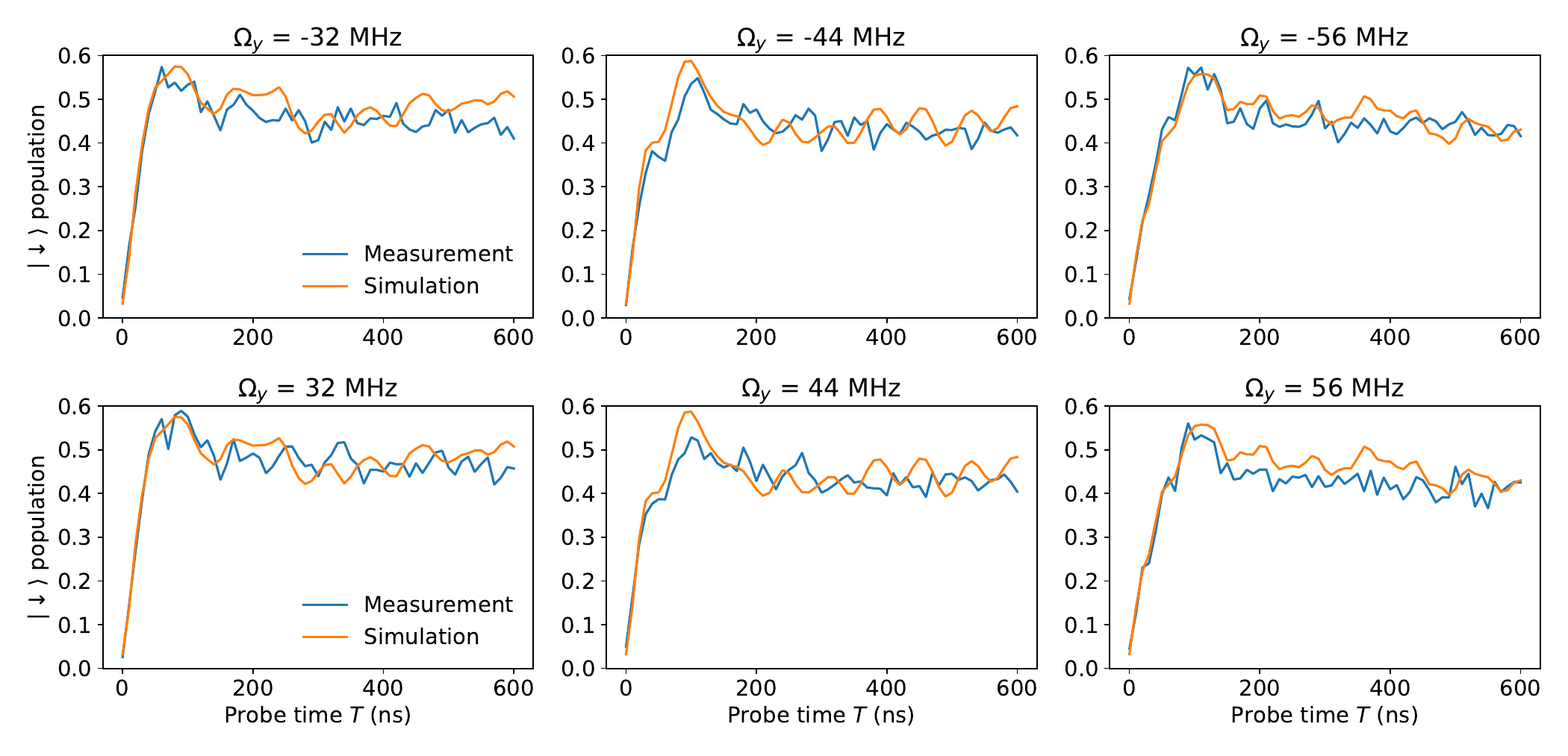}
    \caption{Time dependent NOVEL signals for measurement and simulation of an unpolarized nuclear ensemble. The probe frequencies $\Omega_y$ (panel titles) correspond to the vertical dashed lines in Fig.~\ref{fig:NovelComparison} }
    \label{fig:NovelTimeTraces}
\end{figure*}

To simulate the NOVEL probe, we choose the initial state
\begin{align}
    \hat{\rho}_0 = (F_\text{init}\ketbra{\up}{\up} + (1-F_\text{init})\ketbra{\down}{\down})\otimes \ketbra{\psi_\text{nuc}}{\psi_\text{nuc}}, \label{eq:rho0}
\end{align}
where $F_\text{init}$ is the spin initialisation fidelity (estimated in \ref{sec:SpinInit}) and $\ket{\psi_\text{nuc}}$ is the randomly sampled nuclear state. We then numerically integrate the master equation for the $\pi/2$ pulse, spin locking pulse, and second $\pi/2$ pulse in the NOVEL probe (c.f. Fig. \ref{fig:pulseSequences}e). The final $\ketdown$ electron population is given by $p_\down=\text{Tr}_\text{nuc}\{\bra{\down}\hat{\rho}_1\ket{\down}\}$ where $\hat{\rho}_1$ is the final state and we have traced out the nuclei. This process is repeated for an ensemble of initial $\ket{\psi_\text{nuc}}$.
\\ \\
To simulate an unpolarized ensemble, $\ket{\psi_\text{nuc}}$ samples all three nuclear species from independent thermal distributions following the method in section \ref{sec:thermalSampling}. The assumption of a thermal $^{75}$As ensemble is reasonable, as the $^{75}$As polarisation is classically anticorrelated with the summed gallium ensembles (with near identical hyperfine constants) and thus inherits their thermal characteristics. As we will later discuss, this assumption yields simulations in good agreement with data.
\\ \\
To simulate the ensemble with polarised gallium species, we again sample $^{75}$As from a thermal distribution as we do not expect this distribution to change significantly. 
For $^{71}$Ga, we assume a perfect dark state $\ket{j,-j}$ with $j$ set by the observed magnon Rabi frequency. Meanwhile, the state of $^{69}$Ga is difficult to precisely estimate from measurements. Here, we simply seek to reproduce the main features of the observed $^{69}$Ga resonance, namely its asymmetry, damping, and rise time. We find rough agreement with the experiment by assuming a $^{69}$Ga state $\ket{j,j-\Delta m}$ where $\Delta m\geq 0$ is a dark state deviation which is sampled from an exponential distribution $p_{\Delta m}\propto e^{-\Delta m/\lambda}$. We assume the same degree of polarisation as for $^{71}$Ga, ie. $j/j_0=0.6$. 
$\lambda$ describes the spread in $m$ and we estimate $\lambda=2$ based on the NOVEL signal rise time when driving $^{69}$Ga. Indeed, for an ensemble of identical nuclei, a small $\Delta m$ is necessary to induce asymmetric sidebands following Eq. (\ref{eq:LadderMatrixElement}).
\\ \\
Figure \ref{fig:NovelComparison} shows the measured and simulated NOVEL spectra for an unpolarized and gallium-polarized nuclear ensemble.
We use the unpolarized case to estimate the anisotropy tilt angle $\phi$ which determines the non-collinear hyperfine coupling $a_\perp=a\sin(\phi)$ where $a$ is the single nucleus hyperfine constant. For each nuclear species $i$, we set $a_i=A_i/(N_\text{tot}/2)$ where $A_i$ is the material hyperfine material constant (table \ref{tab:nuclearconstants}) and $N_\text{tot}$ is the effective number of total nuclei estimated in section \ref{sec:numberOfNuclei}. Under the assumption of a thermal ensemble, the activation times of all three species only depend on $\phi$. Figure \ref{fig:NovelComparison} shows excellent agreement between measurement and simulation for $\phi=\SI{0.15}{rad}$ resulting in $a_\perp^{^{71}Ga}=\SI{50}{kHz}$. In both experiment and simulation, the nuclear resonances deviate slightly from the nuclear Larmor frequencies as a result of the spectral overlap between magnon modes. We therefore compare simulation and experiment at the empirically observed resonances. Figure \ref{fig:NovelTimeTraces} shows NOVEL time traces for simulation and experiment further exemplifying their agreement. In experiment and simulation, the dynamics are strongly damped due to the large thermal inhomogeneity of the $|\hat{I}_\pm|$ matrix elements. 
\clearpage
\subsection{Simulating quantum register performance}
\begin{figure*}[h!]%
\centering
\includegraphics[width=1\textwidth]{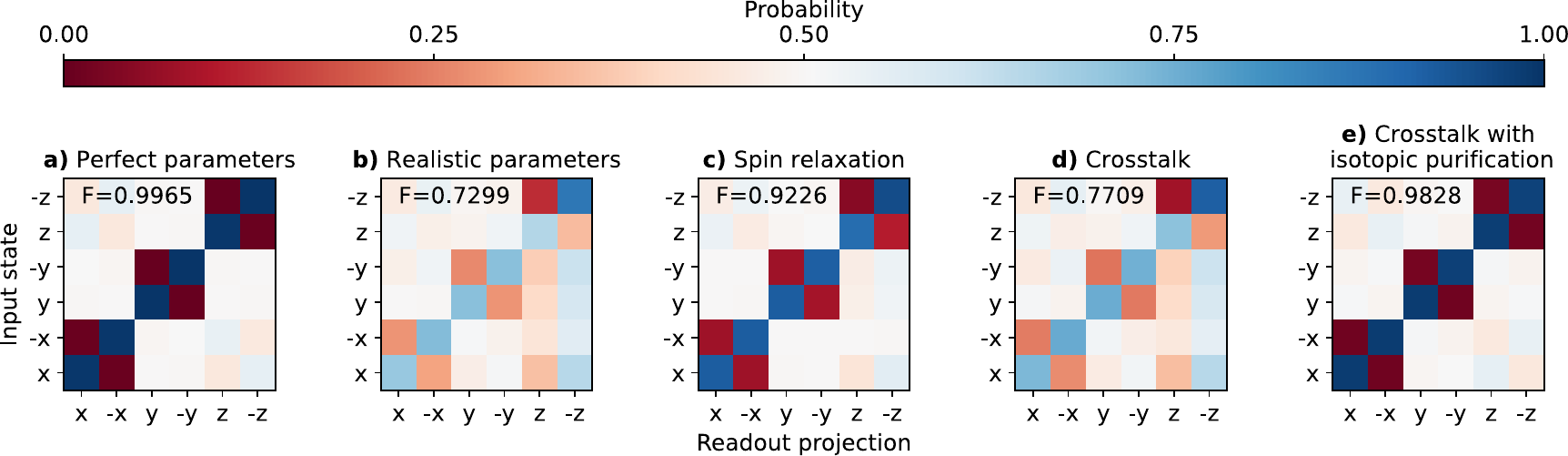}
    \caption{Simulated quantum process tomography (c.f. table \ref{tab:rotations}) of a full prepare-write-read experiment. The simulations represent \textbf{a,} perfect single species nuclear ensemble, \textbf{b,} experimentally studied QD with all imperfections, \textbf{c,} effects of spin relaxation alone in the studied QD, \textbf{d,} effect of nuclear resonance overlap alone in the studied QD, \textbf{e,} effect of nuclear resonance overlap alone in a QD with only $^{75}$As and $^{71}$Ga ensembles prepared in dark states. In \textbf{b,d}, 100 initial nuclear states are randomly sampled.}
    \label{fig:SITomography}
\end{figure*}

To simulate the quantum register, we use the same master equation to simulate all six periods of electron drive in Fig. \ref{fig:pulseSequences}f. The electron reset during storage is incorporated by Krauss operators:
\begin{align}
    \hat{\rho}&\rightarrow \sum_i \hat{K}_i\hat{\rho} \hat{K}_i^\dag,\\
    \hat{K}_1 &= F_\text{init} \ketbra{\up}{\down}\otimes\mathds{I}_\text{nuc},\\
    \hat{K}_2 &= F_\text{init} \ketbra{\up}{\up}\otimes\mathds{I}_\text{nuc},\\
    \hat{K}_3 &= (1-F_\text{init}) \ketbra{\down}{\down}\otimes\mathds{I}_\text{nuc},\\
    \hat{K}_4 &= (1-F_\text{init}) \ketbra{\down}{\up}\otimes\mathds{I}_\text{nuc},\\
\end{align}
where $\hat{K}_3$ and $\hat{K}_4$ represent erroneous electron initialisation.
We repeat the memory experiment for all combinations of input and output states. As our simulation assumes identical nuclei it does not incorporate any dephasing from quadrupolar or Knight field inhomogeneities. We therefore choose to compare fidelity of our simulation and experiment at the shortest storage time.
\\ \\
We first test our simulation using ideal parameters: The couplings to $^{75}$As and $^{69}$Ga are turned off and there are no errors. Fig. \ref{fig:SITomography}a reveals the resulting tomography when using the $^{71}$Ga storage mode. The resulting 0.35\% infidelity stems from the rotating wave approximation not being fully satisfied during state transfer given that the transfer rate $\Omega^+_\text{mag}/2\pi=\SI{3.8}{MHz}$ is not negligible compared to the storage mode frequency $\omega_n/2\pi=\SI{58.4}{MHz}$. This error can be eliminated with pulse shaping or by slower state transfer.

For simulating realistic parameters (Fig. 4b main text), we use the same parameters as for the polarized NOVEL simulation (Fig. \ref{fig:NovelComparison}) and apply a spin-locking drive with $T_\text{sl}=\SI{130}{ns}$ duration and $\Omega_y/2\pi=\SI{56}{MHz}$ Rabi frequency as this results in the maximal signal under NOVEL probing. The simulated tomography is shown in Fig. \ref{fig:SITomography}b.

To estimate the infidelity owing to electron relaxation (Fig. \ref{fig:SITomography}c), we remove the coupling to $^{75}$As and $^{69}$Ga and include the electron relaxation parameterized by $Q=46$ as the only error mechanism. We further numerically optimize $T_\text{sl}$ and $\Omega_y$ which vary from ideal conditions by $<2\%$. 

To estimate the infidelity owing to nuclear resonance overlap (Fig. \ref{fig:SITomography}d), we use the nuclear state estimated from polarized NOVEL but set all other errors to zero.

Finally, we simulate an ideal QD containing only $^{75}$As and $^{71}$Ga ensembles prepared in oppositely polarized $j=0.6j_0$ dark states and with no other errors. In this case, part of the electronic state can get stored in $^{75}$As and be successfully retrieved if the $^{75}$As and $^{71}$Ga modes rephase during storage. This however results in a storage time-dependent fidelity making the fidelity measure ambigious. To circumvent this problem, we reinitialize the $^{75}$As ensemble into its dark state after the first SWAP gate, thereby erasing any information stored. Using the same $\phi=0.15$ as estimated from experiments, we obtain a simulated $F=98.3\%$ and a faster 83.6 ns SWAP gate owing to the increased collective enhancement of the now 100\% abundant $^{71}$Ga ensemble.

\section{Expected inhomogeneous dephasing time of the storage mode}
We estimate the nuclear quadrupolar-induced inhomogeneous dephasing time of the $^{71}$Ga storage mode from NMR measurements of the $\ket{\pm 3/2}\leftrightarrow\ket{\pm 1/2}$ satellite transitions: Ref. \cite{millington-hotze_approaching_2024} reports a $\delta\nu_{69}\sim \SI{7}{kHz}$ FWHM of the $^{69}$Ga satellite transitions. This results in an inhomogeneous dephasing time
\begin{align}
(T_2^*)_{69} = \frac{\sqrt{2}\times\sqrt{8 log(2)}}{2\pi \delta\nu_{69}} = \SI{76}{\micro s},
\end{align}
where the factor $\sqrt{8 log(2)}$ is the FWHM of a Gaussian distribution. We apply a correction to account for the smaller quadrupolar moment of $^{71}$Ga which translates to a proportionally smaller quadrupolar broadening under the assumption that the chemically similar gallium isotopes experience the same distribution of electric field gradients. From the reported ratio of quadropolar moments $Q_{71}/Q_{69}=0.63$ \cite{stone_table_2016}, we estimate a $^{71}$Ga dephasing time of $(T_2^*)_{71}=(T_2^*)_{69}\times (Q_{69}/Q_{71})=\SI{120}{\micro s}$.


\section{Summary of system parameters}
Table \ref{tab:nuclearconstants} enumerates the system parameters recorded in the literature and estimated in this work.

\begin{table}[h]
\begin{tabular}{|l||l|l|l|}
\hline
Species $\alpha$                                                  & $^{75}$As & $^{69}$Ga & $^{71}$Ga \\ \hline \hline
Hyperfine interaction constant $A_\alpha/2\pi$ (GHz)              & 10.4      & 8.7           & 11.1                       \\ \hline
Unit cell concentration $c_\alpha$                                & 1         & 0.604         & 0.396                      \\ \hline
Zeeman splitting $\omega_\alpha/B/2\pi$ (MHz/T)                   & 7.22      & 10.22         & 12.98                      \\ \hline
Zeeman splitting at 4.5 T $\omega_\alpha/2\pi$ (MHz)              & 32.49     & 45.99         & 58.41                      \\ \hline
Effective number of nuclei                                        & 34000     & 20536         & 13464                      \\ \hline
Estimated single nucleus hyperfine  $a_\alpha/2\pi$ (MHz)         & 0.320     & 0.269         & 0.342                      \\ \hline
\end{tabular}
\caption{Physical constants for nuclear species. Concentrations are taken from Ref. \cite{berglund_isotopic_2011}. Hyperfine constants and Zeeman splittings are taken from Ref. \cite{malinowski_notch_2016}. The remaining constants are estimated in this work.}
\label{tab:nuclearconstants}
\end{table}

\bibliography{supplement.bbl}
\newpage

\clearpage